\documentclass[12pt,preprint]{aastex}

\newcommand{\noprint}[1]{}

\shortauthors{Mu\~noz Mar\'{\i}n et al.}
\shorttitle{Near-UV atlas of Seyfert galaxies with ACS-HRC.}

\begin{document}

\title{An Atlas of the circumnuclear regions of 75 Seyfert galaxies in the near-UV with HST Advanced Camera for Surveys.}

\author{
V\'{\i}ctor M. Mu\~noz Mar\'{\i}n\altaffilmark{1}, Rosa M. Gonz\'alez Delgado\altaffilmark{1}, Henrique R. Schmitt\altaffilmark{2}, Roberto Cid Fernandes\altaffilmark{3}, Enrique P\'erez\altaffilmark{1}, Thaisa Storchi-Bergmann\altaffilmark{4}, Tim Heckman\altaffilmark{5}, Claus Leitherer\altaffilmark{6}}
\affil{(1) \em Instituto de Astrof\'{\i}sica de Andaluc\'{\i}a (CSIC), P.O. Box 3004, 18080 Granada, Spain (manuel@iaa.es; rosa@iaa.es)}
\affil{(2) \em Remote Sensing Division, Naval Research Laboratory, Washington, DC; and Interferometrics, Inc., Herdon, VA 20171; henrique.schmitt@nrl.navy.mil.}
\affil{(3) \em Depto. de F\'{\i}sica-CFM, Universidade Federal de Santa Catarina, C.P. 476, 88040-900, Florian\'opolis, SC, Brazil; cid@astro.ufsc.br.}
\affil{(4) \em Instituto de F\'{\i}sica, Universidade Federal do Rio Grande do Sul, C.P. 15001, 91501-970, Porto Alegre, RS, Brazil; thaisa@if.ufrgs.br.}
\affil{(5) \em Department of Physics and Astronomy, Johns Hopkins University, Baltimore, MD 21218.}
\affil{(6) \em Space Telescope Science Institute, 3700 San Martin Drive, Baltimore, MD 21218.}

\begin{abstract}
 We present an atlas of the central regions of 75 Seyfert galaxies imaged in the near-UV with the Advanced Camera for Surveys of the Hubble Space Telescope at an average resolution of $\sim$10\,pc. These data complement archival high resolution data from the Space Telescope at optical and near-IR wavelengths, creating an extremely valuable dataset for astronomers with a broad range of scientific interests. Our goal is to investigate the nature of the near-UV light in these objects, its relation to the circumnuclear starburst phenomenon, and the connection of this to the evolution and growth of the galaxy bulge and central black hole. In this paper, we describe the near-UV morphology of the objects and characterize the near-UV emission. We estimate the size and the luminosity of the emitting regions and extract the luminosity profile. We also determine the presence of unresolved compact nuclei. In addition, the circumnuclear stellar cluster population is identified, and the contribution of the stellar clusters to the total light, at this wavelength, is estimated.
The size of the sample allows us to draw robust statistical conclusions. We find that \mbox{Seyfert 1} galaxies are completely dominated by its bright and compact nucleus, that remains point-like at this resolution, while we find almost no unresolved nucleus in Seyfert 2. The Seyfert types 1 and 2 are quite segregated in an asymmetry vs compactness plot. Stellar clusters are found somewhat more frequently in Sy2 (in $\sim$70\% of the galaxies) than in Sy1 ($\sim$57\%), and contribute more to the total light in Sy2, but this two differences seem to be mostly due to the large contribution of the compact nucleus in Sy1, as the luminosity distribution of the clusters is similar in both Sy types.

\end{abstract}


\section{INTRODUCTION}

\label{sec:Introduction}

The origin of the near-UV continuum in Seyfert 2 nuclei has been a matter of debate for the last decade. In the framework of the unified model (Antonucci 1993), it was first thought that this blue continuum was scattered light from the hidden Seyfert 1 nucleus. Evidence for this picture came mainly from the discovery of broad emission lines in polarized light of Seyfert 2 nuclei (Antonucci \& Miller 1985; Miller \& Goodrich 1990), and high excitation gas extending out from the nucleus with conical or bi-conical morphology (e.g. Wilson, Ward \& Haniff 1988; Tadhunter \& Tsvetanov 1989; P\'erez et al.~1989). However, other optical spectropolarimetry studies (Tran 1995) showed that the polarization of the continuum is lower than that of the broad emission lines, even after the subtraction of an old stellar population typical of the bulge of early type disk galaxies. This result is well understood only if the blue continuum is dominated by another source rather than scattered AGN light. Terlevich, D\'{\i}az \& Terlevich (1990) proposed a stellar origin for this continuum, and Cid Fernandes \& Terlevich (1995) proposed that a heavily-reddened starburst provides this emission. 

The starburst origin for the blue continuum is strongly supported by HST observations. Direct observational evidence that only a small fraction of the total UV light detected in Seyfert 2 galaxies is emitted by a hidden nucleus comes from high resolution UV images of these objects (Heckman et al 1997; Colina et al.~1997; Gonz\'alez Delgado et al.~1998). Powerful circumnuclear starbursts have been unambiguously identified in 40\% of nearby Seyfert 2 galaxies (Gonz\'alez Delgado et al.~2001; Cid Fernandes et al.~2001,\ 2004). These starbursts were originally detected by means of either UV or optical spectroscopy of the central few 100\ pc. Stellar wind absorption lines in the UV spectra (Heckman et al.\ 1997; Gonz\'alez Delgado et al.~1998) and/or high-order Balmer lines of H and HeI, and in some cases Wolf-Rayet features (Gonz\'alez Delgado et al.~1998, 2001; Storchi-Bergmann et al.~2000) show that young and intermediate age stellar population are significant, if not dominant, in the nuclear region ($\sim$100\ pc) of many Seyfert 2. 

On the other hand, it is widely believed that the center of every galaxy contains a super massive black hole (Magorrian et al.~1998), hereafter referred to as SMBH. Strong observational correlations have been observed between the black hole mass and the velocity dispersion of the host bulge (Gebhardt et al.~2000; Ferrarese \& Merritt 2000), suggesting that the formation and growth of the SMBH must be closely linked to the evolution of the bulge itself. In the past, the AGN phenomenon must have coexisted with violent star-formation, but to what extent this is happening today and whether there is a causal connection between them is something that needs to be better understood. The circumnuclear star clusters are very good tracers of this process. This is because the occurrence of star clusters is a common phenomenon in star forming environments as starburst galaxies (Meurer et al.~1995) or at the nuclei of spiral galaxies (Carollo et al.~2002; B\"oker et al.~2002). The nuclear star clusters, also known as stellar nuclei, have been studied spectroscopically in spiral galaxies by Walcher et al.~(2006) and Rossa et al.~(2006). These are massive compact star clusters whose mass seem to be correlated with the luminosity of the host bulge following the same slope than that for SMBH. They are thus intimately linked to the evolution of the galactic bulge. Recent results support the view that SMBH and stellar nuclei have close similarities in their formation and evolution histories (Ferrarese et al.~2006).

The determination of the properties of the nuclear and circumnuclear star cluster population is critical in order to understand the past and present evolution of the bulge and SMBH environment. High resolution imaging combined with a high sensitivity is needed in order to resolve the nuclear star cluster population and disentangle the distribution of extended emission related to the active nucleus and the star forming-regions. In order to do that, we have performed a snapshot survey of a sample of Seyfert galaxies with the Advanced Camera for Surveys (ACS) of the HST in its High Resolution Configuration, with the filter F330W (near UV). This configuration is optimal to detect faint young and middle-aged star-forming regions around these nuclei, and separate their light from the underlying bulge emission. These images complement optical and near-IR images available in the HST archive providing a panchromatic atlas of the inner regions of these objects. These data will allow us to determine the frequency of circumnuclear starbursts, down to levels that cannot be observed from the ground; characterize the properties of these clusters, such as flux, color, size, mass, age, etc.; to study the luminosity function of star clusters and their survival rate close to the AGN; to address questions about the relation between AGNs and starbursts, like the possible connections between the masses of black holes and luminosities of starbursts, and the implications for the evolution of the black holes and their host galaxy bulges.

In this paper we present the Atlas of the observed sample. We have performed a photometric analysis, studied the presence of unresolved nuclei, and carried out a morphological analysis through structure parameters (asymmetry and compactness). We have also estimated the fraction of light coming from star clusters and compact emitting regions. In Section 2 we present the sample, and explain the observations and the data reduction. In Section 3 we explain the analysis process and the results. In Section 4 we present the conclusions of this work. We have included an appendix with a description of the main characteristics of each object.

\section{CATALOGUE, OBSERVATIONS AND DATA REDUCTION}

\label{sec:Sample}

\subsection{Sample Selection}

   We have selected all the Seyfert (Sy) galaxies in the HST archive that had images in these three bands: near ultraviolet with  ACS/HRC F330W, near infrared with NICMOS F160W and optical WFPC2 F606W (in most cases, but also F555W or F547M). In this paper we present an atlas of the near ultraviolet images (ACS/HST F330W) as well as parameters obtained from the analysis of these images. The sample is composed by the galaxies imaged as part of our proposal ID\,9379 (P.I. Schmitt), which is an HST cycle 11 ACS snapshot, plus NGC\,7212 and NGC\,5728 from the proposal ID\,9681 (P.I. Kraemer). These two galaxies have not been imaged with NICMOS, but are included in this paper because they improve our UV study of Seyfert nuclei with their F330W images. The instrumental configuration of the observations is described in more detail in section 2.3. 

The list of objects for proposal ID\,9379 was constructed from the sample presented in Quillen et al.~(2001), consisting in all the Sy observed with NICMOS F160W. Only objects with also WFPC2, most of them in F606W (Malkan et al.~1998), were included in the proposal list. From the original list of 101 objects, 73 were observed during the snapshot, making a total of 75, that will allow us to carry out a statistical study for different types of Sy. From the final sample of 75 objects, 47 (63\%) are classified as Seyfert 2 (Sy2), 14 ($\sim$ 19\%) as intermediate types Sy1.8-1.9, and 14 ($\sim$ 19\%) as Seyfert 1 and Sy1.2-1.5, hereafter refered to as Sy1.

\subsection{Sample Properties}

In Table~\ref{tab:sample_properties} we list the basic properties of the whole sample extracted from NED\footnote{The NASA/IPAC Extragalactic Database operated by NASA/IPAC, Caltech. (\texttt{http://nedwww.ipac.caltech.edu/})}. For the calculation of the distance we have used the Hubble law with H$_0=$75 km\,s$^{-1}$, and the radial velocity data from NED, with the exception of objects with radial velocity $V_r \leq $1200 km\,s$^{-1}$. For those, we used value of the literature, which are: M81 (NGC\,3031) 3.6 Mpc (Freedman 1994); Circinus, 4 Mpc (Freeman 1977) and for the objects:   NGC\,3486 (7.4 Mpc), NGC\,4395 (3.6 Mpc), NGC\,3982 (17 Mpc), NGC\,4258 (6.8 Mpc), NGC\,5005 (21.3 Mpc), NGC\,5033 (18.7 Mpc), NGC\,5194 (7.7 Mpc), NGC\,5273 (21.3 Mpc) and NGC\,6300 (14.3 Mpc) we used the values from Tully (1988). 


\begin{deluxetable}{llcrcccccccccc}
\rotate
\tabletypesize{\tiny}
\tablewidth{0pc}
\tablecaption{Sample Properties}
\tablehead{
  \colhead{Galaxy}   &
  \colhead{Alternative}   &
  \colhead{Spectral} &
  \colhead{Hubble}   &
  \colhead{vel.}  &
  \colhead{Scale}       &
  \colhead{B\_T} &
  \colhead{E(B--V)}  &
  \colhead{axial}  &
  \colhead{L$_{[OIII]}$}  &
  \colhead{Ref.}  &
  \colhead{FIR}  &  
  \colhead{IRAS} &    
  \colhead{IRAS}     
  \\
  \colhead{Name}     &
  \colhead{Name}     &
  \colhead{Class}    &
  \colhead{Type}     &
  \colhead{[km$\,$s$^{-1}$]} &
  \colhead{pc/$^{\prime\prime}$} &
  \colhead{[mag]} &
  \colhead{[mag]} &
 \colhead{ratio (b/a)} &
  \colhead{} &
  \colhead{} &
  \colhead{[10e11 L$_{\odot}$]} &
  \colhead{F$_{12}$/F$_{25}$} &        
  \colhead{F$_{25}$/F$_{60}$}         
\\
  \colhead{(1)}     &
  \colhead{(2)}     &
  \colhead{(3)}    &
  \colhead{(4)}     &
  \colhead{(5)} &
  \colhead{(6)} &
  \colhead{(7)} &
  \colhead{(8)} &
  \colhead{(9)} &
  \colhead{(10)} &
  \colhead{(11)} &
  \colhead{(12)} &
  \colhead{(13)} &        
  \colhead{(14)}         
  }
\startdata

CGCG\,164-019   &                  &  Sy2         & Sa               & 8963  & 579  & 15.3   & 0.026  & 0.875 &   41.4    &  dG92  &  $<$0.672  &    0.735   &    0.436  \\
Circinus        & ESO\,97-G13      &  Sy2         & SA(s)b           & 449   & 19   & 12.1   & 1.455  & 0.435 &   40.21   &  ol94  &    0.135   &    0.275   &    0.275  \\
ESO\,103-G35    &                  &  Sy2         & SA0              & 3983  & 257  & 14.7   & 0.076  & 0.364 &   40.75   &  mw88  &    0.428   &    0.246   &    1.04   \\
ESO\,137-G34    &                  &  Sy2         & SAB(s)0/a?.      & 2747  & 178  & 12.21  & 0.335  & 0.786 &   41.35   &  fe00  &  $<$0.2597 &    0.392   &    0.325  \\
ESO\,138-G1     &                  &  Sy2         & E-S0             & 2740  & 177  & 14.7   & 0.2    & 0.5   &   40.12   &  li88  &  $<$0.186  &    0.273   &    0.664  \\
ESO\,362-G8     &                  &  Sy2         & Sa               & 4785  & 309  & 13.6   & 0.032  & 0.5   &   41.22   &  mu96  &  $<$0.156  & $<$0.316   &    0.297  \\
Fairall49       & IRAS\,18325-5926 &  Sy2         & Sa               & 6065  & 392  & 13.2   & 0.065  & --    &   41.25   &  dG92  &    1.006   &    0.431   &    0.427  \\
IC\,2560        & ESO\,375-G4      &  Sy2         & SB(r)bc          & 2925  & 189  & 12.53  & 0.095  & 0.625 &   40.51   &  gu06  &    0.2097  &    0.437   &    0.246  \\
IC\,4870        & ESO\,105-IG11    &  Sy2-HII     & Pec              & 889   & 57   & 13.89  & 0.113  & 0.563 & --\tablenotemark{a} &    &  $<$0.0073 &     --     &   $<$0.391  \\
IC\,5063        & ESO\,187-G23     &  Sy2         & SA(s)0+          & 3402  & 220  & 12.89  & 0.061  & 0.667 &   41.28   &  sc03  &    0.6409  &    0.302   &    0.642  \\
Mrk\,6          & IC\,450          &  Sy1.5       & SAB0+            & 5640  & 365  & 15.0   & 0.136  & 0.625 &   42.10   &  wh92  &  $<$0.3635 & $<$0.456   &    0.607  \\
Mrk\,40         & Arp151           &  Sy1         &  S0-pec          & 6323  & 409  & 16.8   & 0.014  & 0.429 &   41.18   &  wh92  &     --     &     --     &     --    \\
Mrk\,42         & UGC\,8058        &  Sy1         & SBb              & 7385  & 477  & 15.28  & 0.029  & 0.983 &   40.55   &  wh92  &     --     &     --     &     --    \\
Mrk\,231        &                  &  Sy1         & SA(rs)c?-pec     & 12642 & 817  & 14.41  & 0.010  & 0.769 &   41.91   &  da88  &   29.716   &    0.212   &    0.254  \\
Mrk\,334        & UGC\,6           &  Sy1.8-HII   &  Pec             & 6582  & 425  & 14.38  & 0.047  & 0.7   &   40.254  &  li88  &  $<$1.05   & $<$0.238   &    0.246  \\
Mrk\,461        & UGC\,8718        &  Sy2         & S                & 4856  & 314  & 14.61  & 0.024  & 0.714 &   40.327  &  cg94  &     --     &     --     &     --    \\
Mrk\,471        & UGC\,9214        &  Sy1.8       & SBa              & 10263 & 663  & 14.54  & 0.010  & 0.667 &   40.66   &  da88  &  $<$0.935  &     --     &   $<$0.500  \\
Mrk\,477        &                  &  Sy2         & Compact          & 11310 & 731  & 15.2   & 0.011  & 0.709 &   43.02   &  wh92  &  $<$1.47   & $<$0.463   &    0.4    \\
Mrk\,493        & UGC\,10120       &  Sy1         & SB(r)b           & 9392  & 607  & 14.6   & 0.025  & 0.714 &   40.595  &  li88  &  $<$0.684  & $<$0.926   &    0.422  \\
Mrk\,516        &                  &  Sy1.8       & Sc               & 8519  & 551  & 15.3   & 0.060  & 0.833 &   39.91   &  os81  &  $<$0.915  &     --     &    0.221  \\
Mrk\,915        &                  &  Sy1         & Sb               & 7228  & 467  & 14.82  & 0.063  & 0.833 &   42.07   &  wh92  &  $<$0.569  &    1.625   &    0.711  \\
Mrk\,1210       & UGC\,4203        &  Sy2         & Sa;              & 4046  & 262  & 14.34  & 0.030  & 1.0   &   42.58   &  fa98  &    0.401   &    0.263   &    1.136  \\
NGC\,449        & Mrk\,1           &  Sy2         & (R')S?           &  4780 & 309  & 15.01  & 0.060  & 0.625 &   41.85   &  wh92  &  $<$9.727  & $<$2.24    &    0.348  \\
NGC\,1144       &                  &  Sy2         & S-pec            & 8648  & 559  & 13.78  & 0.072  & 0.636 &   40.256  &  li88  &    2.426   &    0.371   &    0.132  \\
NGC\,1320       & Mrk\,607         &  Sy2         & Sa: sp           & 2663  & 172  & 13.32  & 0.047  & 0.316 &   40.71   &  wh92  &  $<$0.135  &    0.306   &    0.458  \\
NGC\,1672       &                  &  Sy2         & (R'\_1:)SB(r)bc  & 1331  & 86   & 10.28  & 0.023  & 0.833 &   38.53   &  gu06  &    0.5334  &    0.365   &    0.116  \\
NGC\,2639       &                  &  Sy1.9       & (R)SA(r)a:       & 3336  & 216  & 12.56  & 0.024  & 0.611 &   39.45   &  ho97  &  $<$0.212  &     --     &   $<$0.195  \\
NGC\,3031       & M\,81            &  Sy1.8-L     &  SA(s)ab         & -34   & 17   & 7.89   & 0.080  & --    &   39.306  &  li88  &    0.0041  &    0.878   &    0.106  \\
NGC\,3081       &                  &  Sy2         & (R\_1)SAB(r)0/a  &  2385 & 154  & 12.85  & 0.055  & 0.762 &   41.58   &  wh92  &     --     &     --     &     --    \\
NGC\,3227       &                  &  Sy1.5       & SAB(s)-pec    .  &  1157 & 75   & 11.1   & 0.023  & 0.667 &   40.84   &  wh92  &    0.0755  &    0.385   &    0.218  \\
NGC\,3362       &                  &  Sy2         & SABc             &  8290 & 536  & 13.48  & 0.031  & 0.898 &   41.38   &  wh92  &     --     &     --     &     --    \\
NGC\,3393       &                  &  Sy2         & (R')SB(s)ab      &  3750 & 242  & 13.09  & 0.075  & 0.909 &   41.98   &  sc03  &  $<$0.2398 & $<$0.352   &    0.298  \\
NGC\,3486       &                  &  Sy2         & SAB(r)c          &  681  & 36   & 11.05  & 0.022  & 0.732 &   37.96   &  ho97  &  $<$0.0096 & $<$1.115   &    0.0568 \\
NGC\,3516       &                  &  Sy1.5       & (R)SB(s)0        &  2649 & 171  & 12.5   & 0.042  & 0.765 &   41.35   &  wh92  &    0.1223  &    0.489   &    0.529  \\
NGC\,3786       & Mrk\,744         &  Sy1.8       & (R')SAB(r)a-pec  &  2678 & 173  & 13.50  & 0.024  & 0.591 &   40.59   &  wh92  &     --     &     --     &     --    \\
NGC\,3982       &                  &  Sy2         & SAB(r)b          &  1109 & 82   & 11.78  & 0.014  & 0.882 &   40.06   &  wh92  &    0.0719  &    0.571   &    0.1214 \\
NGC\,4253       & Mrk\,766         &  Sy1.5       & (R')SB(s)a:      &  3786 & 245  & 13.70  & 0.020  & 0.8   &   41.77   &  wh92  &    0.3934  &    0.297   &    0.340  \\
NGC\,4258       & M\,106           &  Sy1.9-L     & SAB(s)bc         &  448  & 33   & 9.10   & 0.016  & 0.387 &	  41.02   &  ho97  &     --     &     --     &     --    \\
NGC\,4303       & M\,61            &  Sy2-HII     & SAB(rs)bc        &  1566 & 101  & 10.18  & 0.022  & 0.892 &   40.24   &  li88  &  $<$0.3227 &     --     &   $<$0.0259 \\
NGC\,4395       &                  &  Sy1.8-L     & SA(s)m           &  319  & 17   & 10.64  & 0.017  & 0.833 &	  39.47   &  ho97  &     --     &     --     &     --    \\
NGC\,4565       &                  &  Sy1.9       & SA(s)b? sp       & 1230  & 80   & 10.42  & 0.015  & 0.116 &	  38.71   &  ho97  &  $<$0.848  & $<$1.89    &    0.0778 \\
NGC\,4593       & Mrk\,1330        &  Sy1         & (R)SB(rs)b       & 2698  & 174  & 11.67  & 0.025  & 0.744 &   40.82   &  wh92  &    0.1637  &     --     &   $<$0.327  \\
NGC\,4725       &                  &  Sy2         & SAB(r)ab pec     & 1206  & 78   & 10.11  & 0.012  & 0.710 &	  38.76   &  ho97  &  $<$0.214  &     --     &   $<$0.284  \\
NGC\,4939       &                  &  Sy2         & SA(s)bc          & 3110  & 201  & 11.9   & 0.041  & 0.509 &   41.847  &  li88  &  $<$0.183  &     --     &   $<$0.253  \\
NGC\,4941       &                  &  Sy2         & (R)SAB(r)ab:     & 1108  & 72   & 12.43  & 0.036  & 0.528 &   40.17   &  wh92  &  $<$0.0172 &     --     &   $<$0.425  \\
NGC\,5005       &                  &  Sy2-L       & SAB(rs)bc        &  946  & 103  & 10.61  & 0.014  & 0.483 &	  39.42   &  ho97  &    0.31    &    0.602   &    0.0576 \\
NGC\,5033       &                  &  Sy1.9       & SA(s)c           &  875  & 91   & 10.75  & 0.011  & 0.467 &	  39.36   &  ho97  &    0.1878  &    0.733   &    0.0789 \\
NGC\,5135       &                  &  Sy2         & SB(l)ab          &  4112 & 266  & 12.88  & 0.060  & 0.692 &   41.28   &  wh92  &    1.599   &    0.270   &    0.153  \\
NGC\,5194       & M\,51            &  Sy2-HII     & SA(s)bc pec      &  463  & 37   & 8.96   & 0.035  & 0.616 &   39.14   &  wh92  &    0.0787  &    0.571   &    0.0744 \\
NGC\,5256       & Mrk\,266         &  Sy2         & Pec              &  8353 & 540  & 14.00  & 0.013  & --    &   41.08   &  wh92  &  $<$2.987  & $<$0.471   &    0.140  \\
NGC\,5273       &                  &  Sy1.9       & SA(s)0           &  1064 & 103  & 12..4  & 0.010  & 0.893 &   39.48   &  wh92  &  $<$0.0272 &     --     &   $<$0.417  \\
NGC\,5283       & Mrk\,270         &  Sy2         & S0?              &  3119 & 202  & 14.20  & 0.020  & 0.909 &   41.22   &  wh92  &     --     &     --     &     --    \\
NGC\,5347       &                  &  Sy2         & (R')SB(rs)ab     &  2335 & 151  & 13.4   & 0.021  & 0.765 &   39.96   &  sc03  &    0.8233  &    0.315   &    0.639  \\
NGC\,5548       &                  &  Sy1.5       & (R')SA(s)0/a     &  5149 & 333  & 13.3   & 0.020  & 0.929 &   41.91   &  wh92  &    0.35    &    0.474   &    0.731  \\
NGC\,5674       &                  &  Sy1.9       & SABc             &  7474 & 483  & 13.70  & 0.036  & 0.909 &   41.27   &  gu06  &  $<$0.704  &     --     &   $<$0.24   \\
NGC\,5695       & Mrk\,686         &  Sy2         & SBb              &  4225 & 273  & 13.58  & 0.017  & 0.715 &   41.09   &  wh92  &  $<$0.152  &     --     &   $<$0.525  \\
NGC\,5728       &                  &  Sy2         & (R\_1)SAB(r)a    &  2788 & 180  & 12.81  & 0.101  & 0.581 &   41.526  &  li88  &  $<$0.352  & $<$0.395   &    0.096  \\
NGC\,5940       &                  &  Sy1         & SBab             & 10172 & 658  & 14.32  & 0.041  & 1.0   &   41.30   &  wh92  &  $<$0.88   &     --     &   $<$0.316  \\
NGC\,6300       &                  &  Sy2         & SB(rs)b          & 1109  & 69   & 10.98  & 0.097  & 0.667 &   39.84   &  sp89  &    0.1112  &    0.344   &    0.153  \\
NGC\,6814       &                  &  Sy1.5       & SAB(rs)bc        & 1563  & 101  & 12.06  & 0.183  & 0.933 &   40.26   &  wh92  &    0.0985  &    0.559   &    0.104  \\
NGC\,6951       &                  &  Sy2-L       & SAB(rs)bc        & 1424  & 92   & 11.64  & 0.366  & 0.564 &	  38.99   &  ho97  &    0.170   &    0.385   &    0.0867 \\
NGC\,7130       &  IC\,5135        &  Sy2-L       & Sa pec           & 4842  & 313  & 12.98  & 0.029  & 0.933 &   41.27   &  sp90  &    2.083   &    0.294   &    0.128 \\
NGC\,7212       &                  &  Sy2         & Sab              & 7984  & 516  & 14.78  & 0.072  & --    &   42.34   &  wh92  &  $<$1.35   & $<$0.466   &    0.245 \\
NGC\,7319       &                  &  Sy2         & SB(s)bc pec      & 6747  & 436  & 14.11  & 0.079  & 0.765 &   41.17   &  wh92  &     --     &     --     &     --   \\
NGC\,7469       &                  &  Sy1.2       & (R')SAB(rs)a     & 4892  & 316  & 13.0   & 0.069  & 0.733 &   41.84   &  wh92  &    3.599   &    0.237   &    0.203 \\
NGC\,7479       &                  &  Sy2-L       & SB(s)c           & 2381  & 154  & 11.60  & 0.112  & 0.756 &   38.44   &  dG92  &    0.474   &    0.226   &    0.274 \\
NGC\,7496       &                  &  Sy2         & (R')SB(rs)bc     & 1649  & 107  & 11.91  & 0.010  & 0.909 &   39.60   &  gu06  &    0.134   &    0.178   &    0.178 \\
NGC\,7674       & Mrk\,533         &  Sy2-HII     & SA(r)bc pec      & 8671  & 560  & 13.92  & 0.059  & 0.909 &   42.26   &  wh92  &    3.188   &    0.375   &    0.345 \\
NGC\,7743       &                  &  Sy2         & (R)SB(s)0+       & 1710  & 111  & 12.38  & 0.070  & 0.867 &	  39.60   &  ho97  &  $<$0.031  &     --     &   $<$0.433 \\
UGC\,1214       & Mrk\,573         &  Sy2         & (R)SAB(rs)0+     & 5174  & 334  & 13.68  & 0.023  & 1.0   &   42.30   &  wh92  &  $<$0.3358 & $<$0.363   &    0.630 \\
UGC\,1395       &                  &  Sy1.9       & SA(rs)b          & 5208  & 337  & 14.18  & 0.075  & 0.769 &   40.89   &  wh92  &  $<$0.308  &     --     &   $<$1.8   \\
UGC\,2456       & Mrk\,1066        &  Sy2         & (R)SB(s)0+       & 3605  & 233  & 13.64  & 0.132  & 0.588 &   41.20   &  wh92  &    0.763   &    0.216   &    0.221 \\
UGC\,6100       &                  &  Sy2         & Sa?              & 8844  & 572  & 14.30  & 0.012  & 0.617 &   41.53   &  sc03  &  $<$0.623  &     --     &   $<$0.426 \\
UGC\,12138      &                  &  Sy1.8       & SBa              & 7487  & 484  & 14.24  & 0.085  & 0.875 &   41.40   &  sc03  &  $<$0.633  &     --     &    0.477 \\
UM\,625         &                  &  Sy2         & S0               & 7492  & 484  & 17.43  & 0.062  & 0.848 &   41.48   &  Te91  &     --     &     --     &     --  \\%
\enddata
\label{tab:sample_properties}
\tablecomments{Col.\ (1): Galaxy name. Col.\ (2): Alternative name. Col.\ (3): Spectral class. (LINER = L). Col.\ (4): Hubble type. Col.\ (5): Radial velocity.
Col.\ (6):  Angular scale calculated from the distance. Col.\ (7): Total asymptotic magnitude in B, B\_T, from RC3 catalogue. Col.\ (8): Reddening, E(B-V).
  Col.\ (9): Axial ratio (b/a). (All these quantities, but the scale, were extracted from NED.) Col.\ (10): Logarithm of [OIII]$\lambda$5007 luminosity in units erg/s. 
Col.\ (11): References for column (10); 
cg94: Cruz-Gonz\'alez et al.~(1994); 
da88: Dahari \& De Robertis (1998); 
dG92: de Grijp et al.~(1992); 
fa98: Falcke, Wilson, \& Simpson (1998); 
fe00: Ferruit, Wilson, \& Mulchaey (2000); 
gu06: Gu et al.~(2006); 
ho97: Ho, Filippenko \& Sargent (1997); 
li88: Lipovetsky, Neizvestny, \& Neizvestnaya (1988);
mw88: Morris \& Wald (1988); 
mu96: Mulchaey, Wilson, \& Tsvetanov (1996); 
ol94: Oliva et al.~(1994); 
os81: Osterbrock (1981); 
sc03: Schmitt et al.~(2003); 
sp89: Storchi-Bergmann \& Pastoriza (1989); 
st90: Storchi-Bergmann, Bica, \& Pastoriza (1990); 
te91: Terlevich et al.~(1991). 
wh92: Whittle (1992); 
Col.\ (12): IR luminosity from IRAS fluxes calculated with the formula from Sanders \& Mirabel (1996). Col.\ (13) and (14): IRAS flux ratios, F$_{12}$/F$_{25}$ and F$_{25}$/F$_{60}$.
}  
\tablenotetext{a}{IC4870 is a Wolf-Rayet galaxy}
\end{deluxetable}

\begin{figure}
\plotone{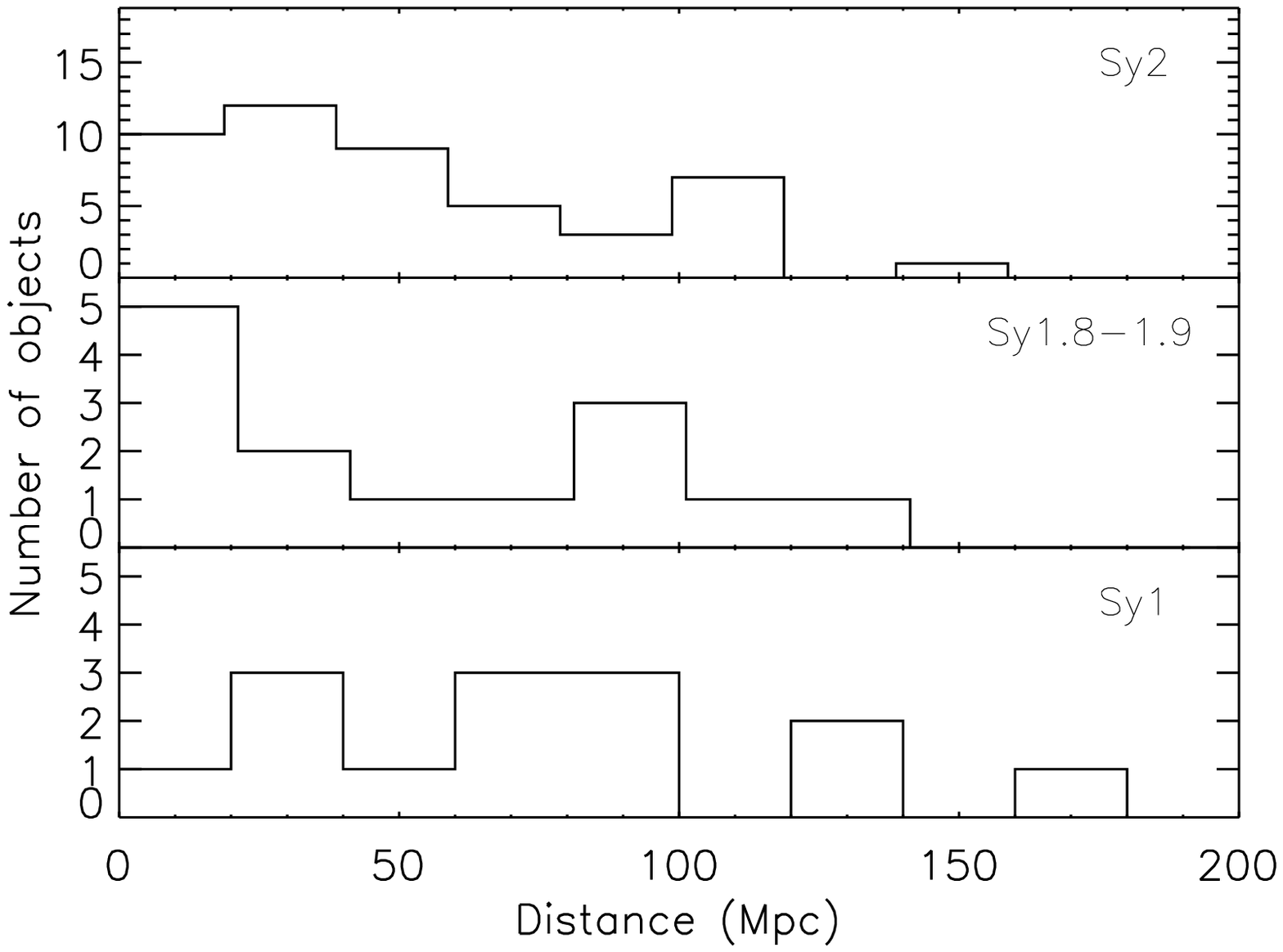}
\epsscale{0e}
\caption{The upper, middle and lower panel show the distribution of the distance for the Sy2, Sy1.8-1.9, and Sy1 galaxies, respectively.}
\label{fig:f1}
\end{figure}

   In order to understand the possible biases of our sample we compare the general properties of our galaxies with those in two bona-fide samples of Sy galaxies in the literature, the CfA and RSA Seyfert subsamples. From the 48 Sy galaxies in the CfA catalogue presented in Huchra \& Burg (1992) (see also McLeod \& Rieke 1995), 24 are in our sample as well. On the other hand, 38 out of 75 of our galaxies belong to the extended RSA Sy sample compiled by Maiolino \& Rieke (1995) (sample D in their work). We thus explore 50\% of CfA and 42\% of RSA, with only 10 of our galaxies occurring in both of them. As CfA has been shown to lack some bright Seyferts, some of which are in our sample, we have used instead the extension to the CfA sample presented in Alonso-Herrero et al.~(1993), in which they add a total of nine galaxies previously classified as LINER. From now on we will name these subsamples CfA and RSA respectively. The mean distance, d, of our sample is in between both comparison ones, as $<$d$_{RSA}>=34$ Mpc and $<$d$_{CfA}>$ is three times larger than this (Maiolino \& Rieke 1995), while the mean distance of our sample is $<$d$>=57$ Mpc. As in most of the Sy samples in the literature, in our sample Sy1 are, on average, more distant than Sy2. The origin of this bias is that the luminosity of the nucleus compared to the host bulge luminosity is smaller in Sy2 than in Sy1 AGNs. Still, the distance distribution of the galaxies looks homogeneus up to 100 Mpc (Fig.~\ref{fig:f1}). The standard deviation of the distribution is 41 Mpc, indicating that the range of distances is quite large. With ACS we are able to achieve a much better resolution than any ground-based study. The scale of our images ranges from less than 1 pc\,pixel$^{-1}$ for the nearest objects to about 20 pc\,pixel$^{-1}$ for the furthest, with a mean value of 6 pc\,pixel$^{-1}$.

\begin{figure}
\epsscale{0.9}
\plotone{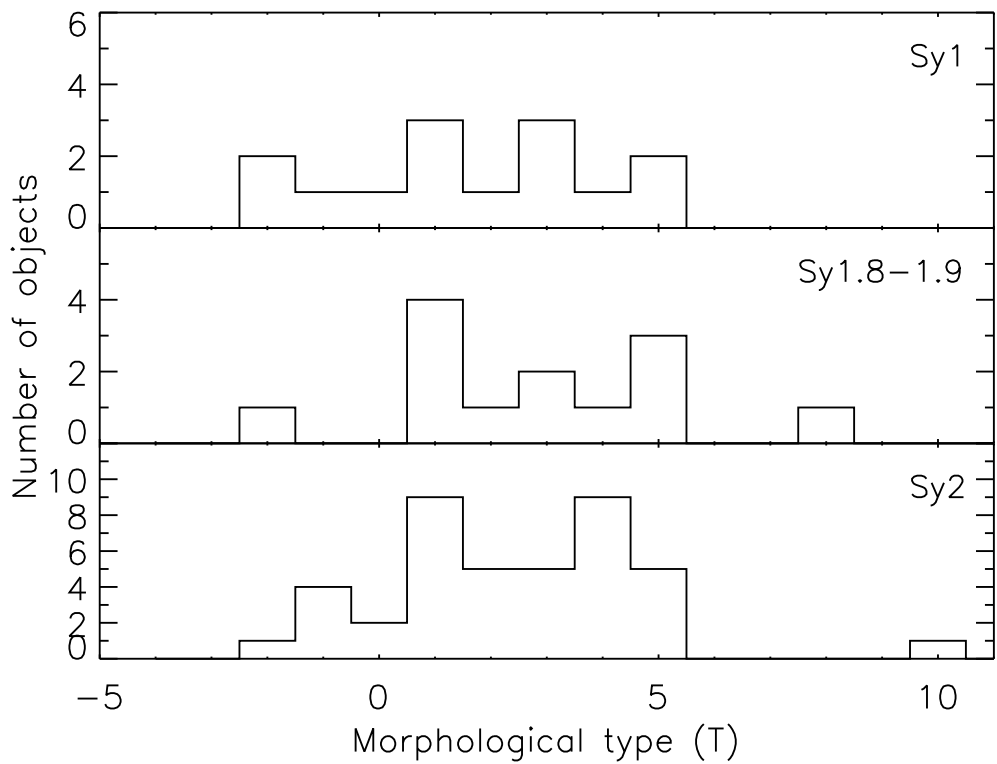}
\epsscale{0.9}
\caption{Distribution of Hubble types for the different subsamples of Sy activity class. The de Vaucouleurs classification (T) is used: e.g. S0=--1, S0a=0, Sa=1, Sab=2, Sb=3, etc. The T $<$ --1 stand for ellipticals. The three subsamples are equivalent on average and median values.}
\label{fig:f2}
\end{figure}

\begin{figure}
\plotone{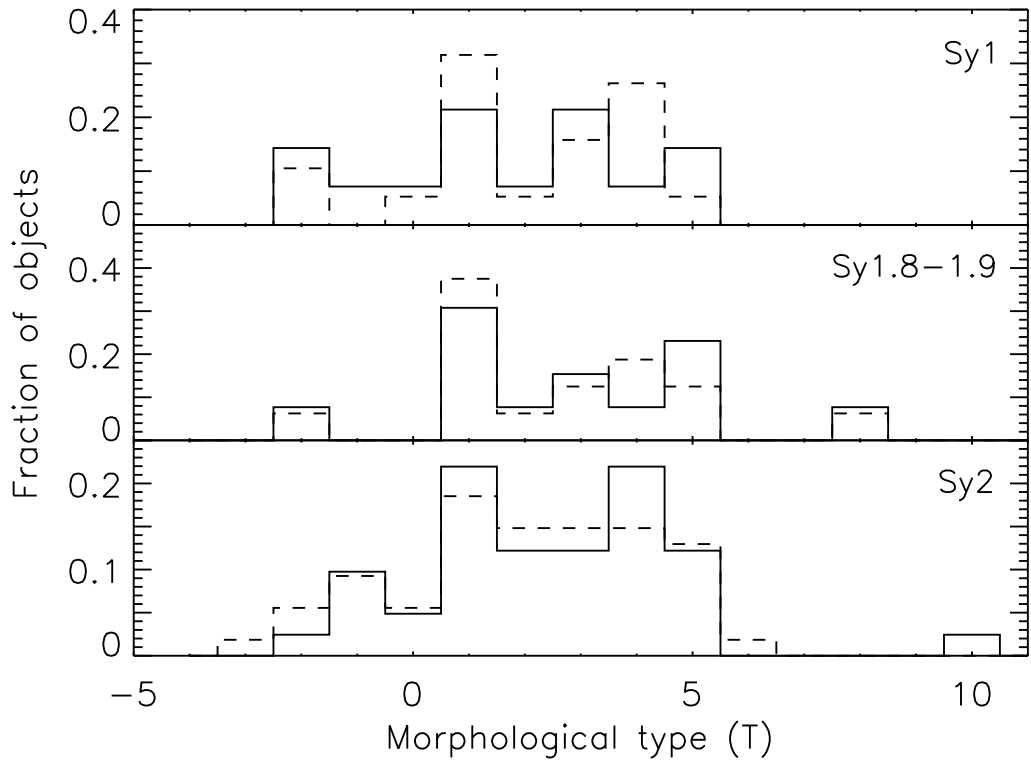}
\epsscale{0.9}
\caption{Comparison between the distribution of morphological types of our sample (full lines) and the RSA sample (dashed lines). The histograms do not differ much. On average both samples are equivalent (see discussion in text).}
\label{fig:f3}
\end{figure}

\begin{figure}
\plotone{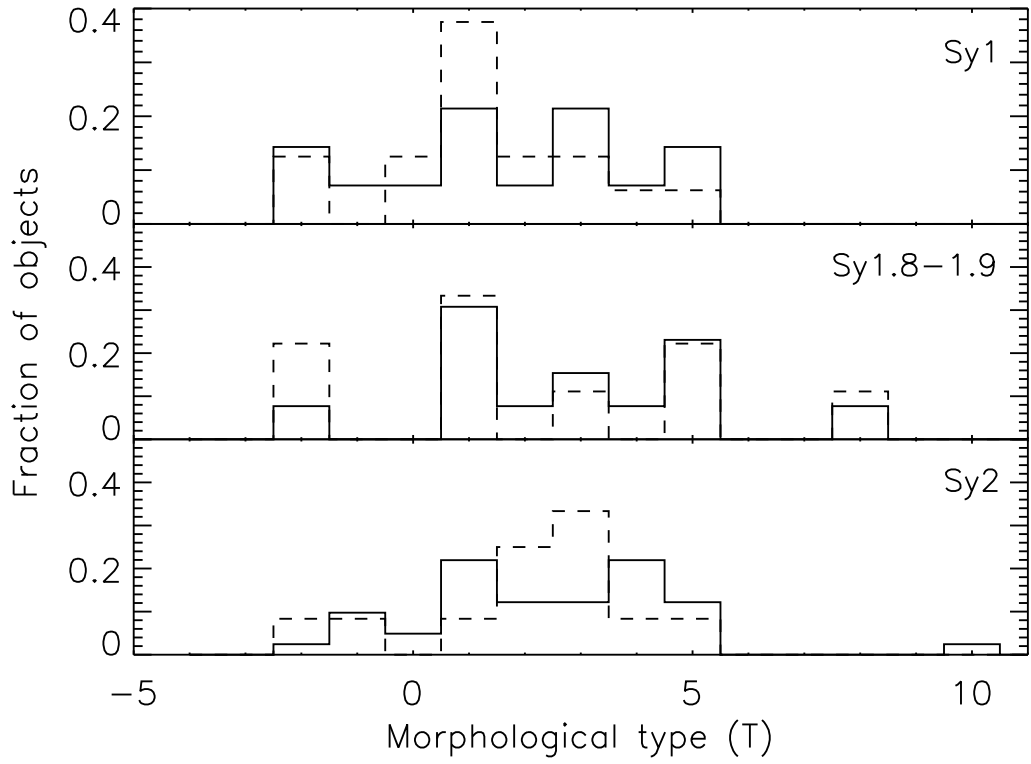}
\epsscale{0.9}
\caption{Same comparison as in Fig.~\ref{fig:f3} for our sample (full lines) and CfA (dashed). The two distributions are very similar as well.}
\label{fig:f4}
\end{figure}

We have checked for possible bias in the distribution of Hubble types among different types of Sy. This is plotted in Fig.~\ref{fig:f2}. Peculiar galaxies and those classified as uncertain, are excluded. As it is widely known, Seyfert nuclei are found mostly in spiral galaxies, with preference for early types (see e.g. Moles, M\'arquez \& P\'erez 1995, and references therein). The histograms do not seem to differ much. In Figs.~\ref{fig:f3} and \ref{fig:f4} we have plotted a comparison of CfA and RSA samples, finding a good agreement between our sample and the ones we have used for comparison. In order to quantify both statements above, we used the de Vaucouleurs classification from RC3 catalogue (T), that is S0$=-1$, S0a$=0$, Sa$=1$, Sab$=2$, etc. The resulting mean, median and standard deviation of the spiral types for our sample and the comparison ones are summarised in Table~\ref{tab:morphology}. Our three subsamples of Seyfert activity show quite similar values of mean and median T. This is enough to ensure that the differences we find among groups of activity type do not arise from differences in the Hubble morphology. Applying the same classification to the other samples we get a good matching with our own sample, although CfA galaxies tend to be of a bit earlier type. 


\begin{deluxetable}{lccccccccc}
\tabletypesize{\small}
\tablewidth{0pc}
\tablecaption{Statistics of Hubble Type.}
\tablehead{
  \colhead{ }   &
  \colhead{ }   &
  \colhead{\textbf{Sample}}  &
  \colhead{ }   &
  \colhead{ }   &
  \colhead{\textbf{RSA}}   &
  \colhead{ }   &
  \colhead{ }   &
  \colhead{\textbf{CfA}}  & 
  \colhead{ }   
\\
  \colhead{ }   &
  \colhead{   mean} &
  \colhead{   median} &
  \colhead{   $\sigma$}  &
  \colhead{   mean} &
  \colhead{   median} &
  \colhead{   $\sigma$}  &
  \colhead{   mean} &
  \colhead{   median} &
  \colhead{   $\sigma$}  
  }
\startdata
\textbf{Sy1}       &  1.6 & 2 & 2.3  & 2.0  &  2  &  2.0  & 1.3  &  1  &  1.9  \\
\textbf{Sy1.8-1.9} &  2.8 & 3 & 2.6  & 2.8  &  3  &  2.6  & 2.2  &  1  &  3.3  \\
\textbf{Sy2}       &  2.4 & 2 & 2.3  & 2.0  &  2  &  2.2  & 2.1  &  3  &  2.0  
\enddata
\tablecomments{The table shows the mean, median and standard deviation of Hubble type for each subsample, following a de Vaucouleurs classification. All the subsamples are represented, on the mean, by early type spirals (Sa=1, Sab=2, Sb=3, etc.)}
\label{tab:morphology}
\end{deluxetable}


\begin{figure}
\plotone{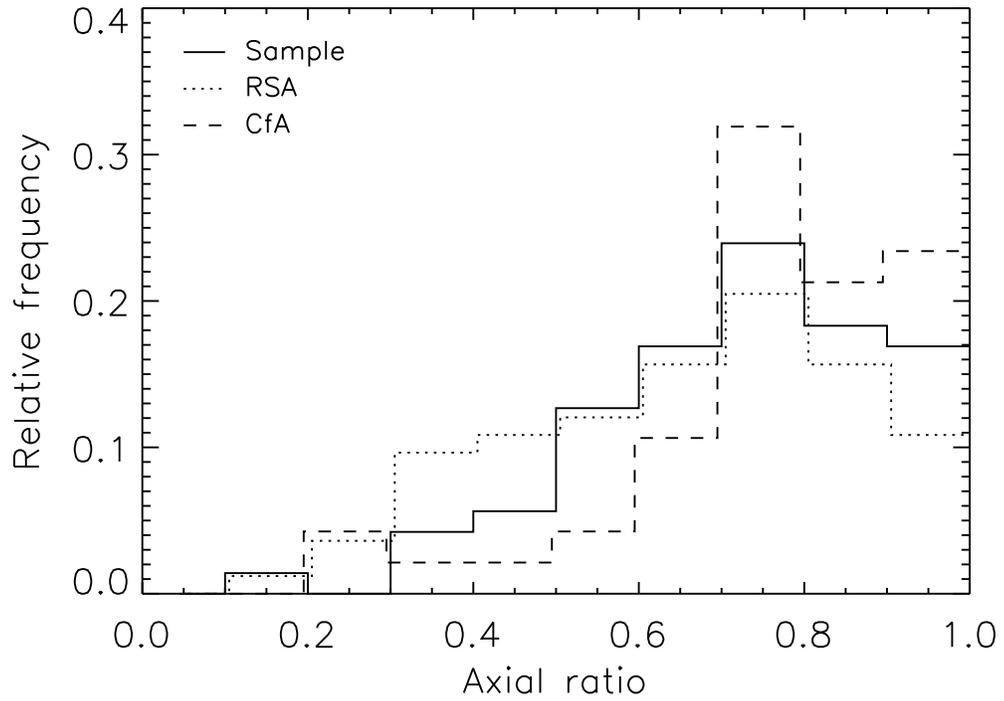}
\epsscale{0.9}
\caption{Compared histogram of the axial ratio (b/a) of the galaxies of the three different samples. Numbers are relative to the total number of objects in each sample. Our sample seems to lay in between the CfA and the RSA, and it is less biased against edge-on galaxies than CfA.
}
\label{fig:f5}
\end{figure}

Finally, we have obtained the distribution of the axial ratio, that is, the minor over the major axis of the galaxy (b/a). This gives an idea of the inclination angle of the galaxy, with high values for objects seen nearly face-on, and low values for edge-on ones. Due to internal galactic absorption more objects with high b/a are expected, leading to a power law distribution for a magnitude-limited sample (Maiolino \& Rieke 1995). Thus, most samples, including ours, are biased against edge-on galaxies. In Fig.~\ref{fig:f5} we plot the distribution of the axial ratio for the three samples. Ours is in between the less biased RSA and the CfA (more affected by this effect). This happens naturally because our sample has a mean distance in between the other two. Maiolino \& Rieke (1995) find that Sy1 tend to occur more often in face-on than the intermediate type Sy. Within our data also a slight trend in the distribution of Seyfert activity with inclination is observed (Fig.~\ref{fig:f6}). This effect is not very severe, as the median value of b/a does not change much among groups, with 0.85 for Sy1, 0.75 for Sy1.8-1.9 and 0.75 for Sy2 galaxies.

\begin{figure}
\plotone{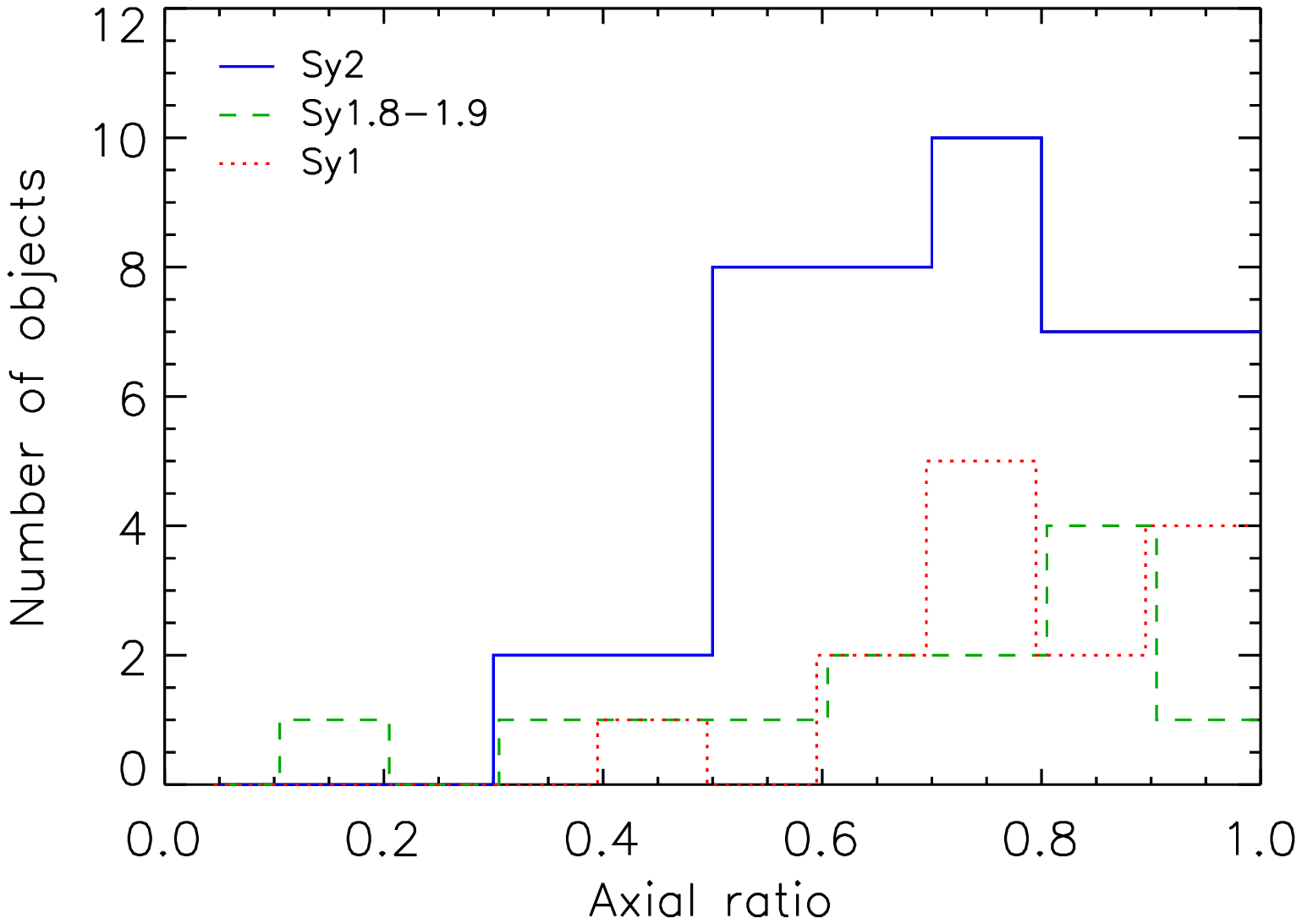}
\epsscale{0.9}
\caption{Distribution of the axial ratio (b/a) for the three activity classes. The typical trend of less objects at low inclinations is seen for this sample, although no clear trend with Sy type is observed.}
\label{fig:f6}
\end{figure}

\subsection{Observations and Data Reduction}
The sample was imaged with the Advanced Camera for Surveys (ACS) in its high resolution configuration (HRC), which provides a pixel size of 0.027\,arcsec. The filter chosen is F330W, that from the UV filters of ACS, has the highest throughput, negligible red leak and minimal contamination by line emission. A higher red leak, as that from F220W, would result in a higher background level with lower S/N detection for young star clusters. The filter F330W has a bandwidth of $\sim400$ \AA \,centred around 3300 \AA, therefore the only strong emission lines contributing to this filter are [NeV]$\lambda$$\lambda$3346,3426. This type of emission will be normally extended, thus is not a problem for measuring compact objects such as clusters (see section 4). 

To allow an easier removal of cosmic rays (CRs) two exposures of 10 min each where made for a total of 1200 s. For 19 of the brightest galaxies the exposure was further split in 1140 s and 60 s exposures, in order to be able to study the possible saturated core. In these cases we have worked with the longest exposure when possible. The only exceptions are NGC\,7212 and NGC\,5728, that have 2550 s exposure.

We downloaded the images from the HST archive\footnote{\texttt {http://archive.stsci.edu/hst/}} with the calibration `on the fly' option, that corrects the images from bias, dark and flat-field subtraction with the most up-to-date ACS reference files and bad pixel tables. Also CRs are rejected and the exposures combined in a single image. The ACS field of view is heavily distorted, due to the design with a minimum number of components and a significant tilt of the detector. When projected in the sky plane the square detector becomes rhombus-shaped. One of the automatic tasks of the ACS pipeline performs the necessary distortion correction. The final products are both astrometrically and photometrically accurate.

The pipeline CRs removal task uses very mild values for the parameters, avoiding to remove by mistake the centre of bright stars or galactic nuclei. We found that many of the images had conspicuous artifacts, such as CRs that were not removed by the pipeline. We tried out several IRAF tasks for the identification and removal of the CRs. Due to the small FWHM of the PSF (1.8 pix) and the low S/N in some of the images we could not get a satisfactory result with these routines. Finally we had to remove the artifacts by hand from the region of interest, after comparing the images with those from WFPC2 to help us to discern the CRs from small star clusters.

The background for every object was determined with the IRAF\footnote{IRAF (Image Reduction and Analysis Facility) is distributed by the National Optical Astronomy Observatories, which are operated by AURA, Inc., under cooperative agreement with the National Science Foundation.} task FITSKY. The mean value and the standard deviation of the background ($\sigma_s$) was measured from several apertures in the outer regions of the image. The values of $\sigma_s$ have a very low scatter (0.004-0.005 counts\,s$^{-1}$) due to the constant instrumental configuration and the similar exposure time in different images. The larger scatter of the background values (from 0.0015 to 0.005 counts\,s$^{-1}$, equivalent to 21--22.5 mag/arcsec$^{2}$) suggests that in some cases this is not a real sky determination, but background light from the galaxy itself. As an example of this there are some cases in which the galaxy fills the field of view (e.g. M81).

When imaging very bright objects, as some Seyfert nuclei, the possible effects of saturation have to be accounted for. The detector can reach physical saturation when more charge is released in a single pixel than what it can accumulate, resulting in charge being spilled to adjacent pixels and some flux getting lost. In particular, the default gain value of the HRC chip falls short of sampling the full well depth ($\sim$165,000 e$^{-}$) by some 22\%. There are 11 objects that overcome this threshold. The core-saturated objects are: Mrk\,231, NGC\,3227, NGC\,3516, NGC\,4593, NGC\,5548, NGC\,7469, UGC\,12138 (with a 60s image), and Mrk\,493, Mrk\,915, NGC\,5273, NGC\,6814 (without a 60s image). In the case of the objects with a 60s image the correction to be made can be calculated by comparing the photometry between the long and short exposures. See below for further discussion.


\section{RESULTS}

Figs.~7.1--7.75 show the central emission of all the galaxies in the sample. We have chosen the field of view and scaling in order to enhance the most interesting features\footnote{All the figures in figure sets 7, 8, 9, and 20, can be downloaded in EPS file format in the URL: http://www.iaa.es/$\sim$manuel/publications/paper01.html}. In addition, for some galaxies we show a nuclear close-up in Figs.~8.1--8.12. We have divided the objects in Seyfert types, in order to better appreciate the common characteristics of each type. In the Appendix we describe the main morphological components of these galaxies, and also we show figures of all the objects showing the whole HRC field of view. The morphology of the objects is as irregular as varied. There are many different features within the sample: star-forming rings, spirals, clumpy diffuse light emission, plain PSF-dominated objects, complete lack of compact nucleus, etc. From Figs.~7.1--7.14 it can be seen that every Sy1-1.5 possess a bright star-like nucleus, which precludes the observation of the inner morphology. In several cases regions of star-formation and rings can be seen in the images as well. The morphology of intermediate type Sy (Figs.~7.15--7.28) is more varied. Some objects have a compact nucleus. The morphology can be clumpy or diffuse, and some objects show dust absorption features or ionization cones. 
For the Sy2 galaxies (Figs.~7.29--7.75) the morphology is mostly clumpy, with frequent star-formation regions. These are often arranged in rings or spiral arms. There are some objects showing instead a biconical or symmetrical structure as ionization cones. 
When a very bright nucleus is present, as in the images of galaxies from Sy1 to Sy1.9, some artifacts may appear, such as inner rings or clumps very close to the nucleus. These are caused by the instrumental PSF, that shows not only the diffraction spikes, but also clumpy ring-like wings that can be confused with actual star-forming rings (e.g. see image of Mrk\,42, Mrk\,493 or Mrk\,915).

\begin{figure}
\figurenum{7}
\addtocounter{figure}{1}
\includegraphics[angle=270,width=0.9\textwidth]{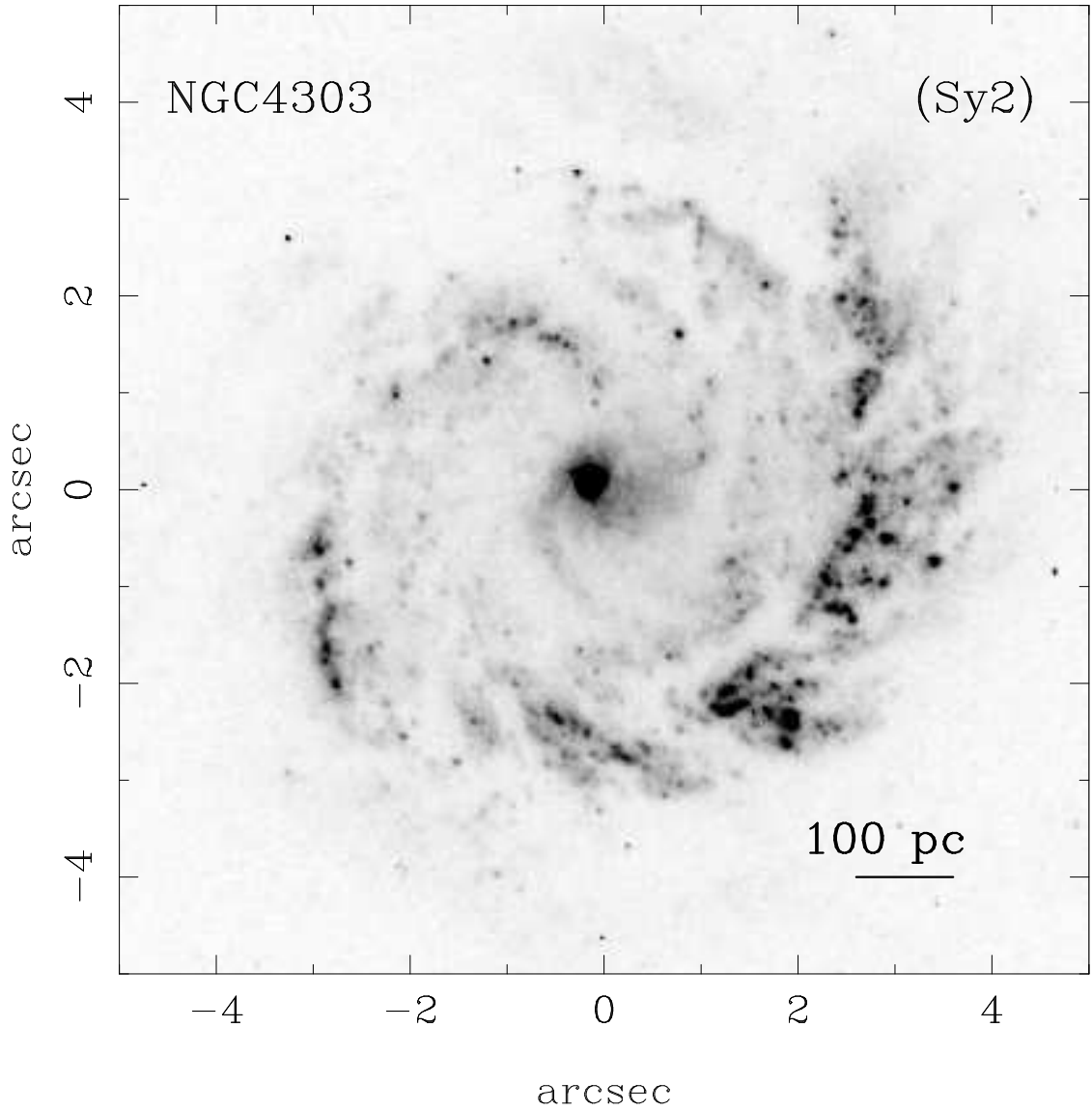}
\caption{Example of a near-UV ACS image. The complete set of images for the whole sample are available in the electronic version of the Journal. The field of view and contrast is chosen to show the most interesting parts and structure of each object. North is up, east to the left.}
\end{figure}

\begin{figure}
\figurenum{8}
\addtocounter{figure}{1}
\includegraphics[angle=270,width=0.9\textwidth]{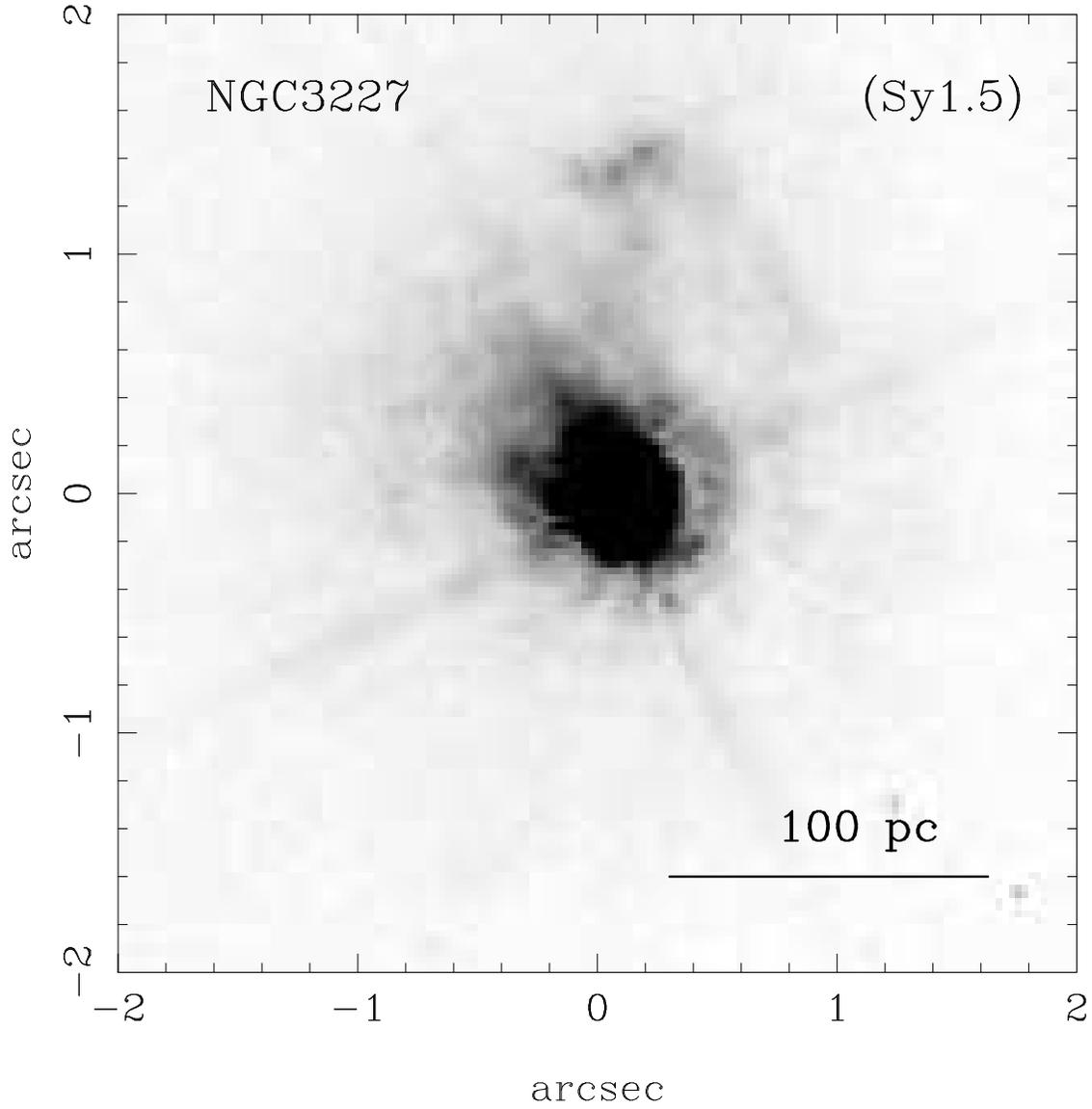}
\caption{Close-up of some galaxies with interesting nuclear structure or a large luminosity range within the image. The complete set of 12 panels is available in the electronic edition of the Journal.}
\end{figure}



\subsection{Profiles}

Due to the irregularity of the isophotes, the common method of elliptical isophotal fitting was ruled out. Instead, we performed a photometric analysis using circular apertures. Aperture photometry was carried out with IRAF task PHOT in order to determine the surface brightness profile and the luminosity and magnitude curves of growth. The surface brightness profile was computed by measuring the mean number of counts in circular annuli.
First of all, an accurate position of the nucleus was computed when a compact source was clearly detected in the images. All the apertures were centered in the position determined by the centroid of the compact nuclear source. In the cases in which the nucleus was too obscured an approximate position was estimated by taking as a reference the nuclear position in the WFPC2 optical images (F606W and F814W), as these wavelengths are less affected by dust extinction, and using the galaxy features visible in both bands to align them. For these objects (Circinus, ESO\,137-G34, ESO\,362-G8, NGC\,1672, NGC\,2639, NGC\,5194, NGC\,5256, NGC\,5728, NGC\,6300 and NGC\,6951) a precision of a few ACS pixels was achieved and that was enough for the rest of the analysis, so the inner regions are qualitatively described by the profiles.

The Point Spread Function (PSF) of the instrumental configuration was computed with the software TinyTim\footnote{http://www.stsci.edu/software/tinytim/tinytim.html}. The resulting PSF was compared with the radial profiles of several isolated stars in some images, showing a very good agreement. We then compared the PSF with the surface brightness profiles in order to determine whether or not the galaxies show a compact resolved nucleus. The PSF of the ACS-HRC at this wavelength has 1.8 pixels full-width at half maximum (FWHM), so a compact source with a FWHM larger than 0.05$^{\prime\prime}$ should be seen slightly extended. The occurrence of a nuclear point source was determined by eye inspection of the radial profiles over-plotted, in a logarithmic scale, to the PSF profile and normalized to the same peak value. The results are summarized in Table~\ref{tab:nuc}. This table shows the number of objects of each Sy type for which the ACS-HRC shows a compact and unresolved nucleus. We have not found resolved nuclei in any galaxy from Sy1 to Sy1.5 type. On the other hand, most Sy2 nuclei appear resolved or absent (heavily obscured). For the intermediate types 1.8--1.9 the situation is something in between, with approximately one third of the nuclei resolved, one third remaining point-like, and the rest being difficult to discern. The objects with nucleus at the limit of resolution are Mrk\,516, Mrk\,334, NGC\,4565, UGC\,1395, and the Sy2 CGCG\,164-019. This result is similar to that obtained at other wavelengths; for example, Nelson et al.~(1996) show that in the red, Sy2 in general lack a compact nucleus while Sy1-1.5 are dominated by a bright unresolved nuclear source.

\begin{deluxetable}{lccccc}
\tabletypesize{\small}
\tablewidth{0pc}
\tablecaption{Frequency of point-like nuclei}
\tablehead{
  \colhead{\textbf{Sy type}}   &
  \colhead{\textbf{No nucleus or resolved}}   &
  \colhead{\textbf{Unresolved}}  &
  \colhead{\textbf{At resolution limit}}   &
  \colhead{\textbf{Total}}   
  }
\startdata
\textbf{Sy 2} & 43 (91.5\%) & 3 (6.4\%) & 1 (2.1\%) & 47\\
\textbf{Sy 1.8--1.9} & 5 (35.7\%)& 5 (35.7\%) & 4 (28.6\%) & 14\\
\textbf{Sy 1--1.5} & 0 & 14 (100\%) & 0 & 14\\
\textbf{Total} & 48 & 22 & 5 & 75
\enddata
\tablecomments{The table shows the number of objects of each type for which we resolve the nucleus. The numbers in brackets are the equivalent percentages relative to the total number of objects of each type. Note that no Sy1 nucleus is resolved.}
\label{tab:nuc}
\end{deluxetable}

The computed profiles are shown in Figs.~9.1-9.75. By inspection of these plots we have classified the surface brightness profiles in one of these three categories:
\begin{itemize} 

\item{\bf Exponential:} The differential flux has an exponential dependence with the radius, so the dependence of the surface brightness ($\mu$) is linear with r. This is the classical  model to fit well the surface brightness profile of some dwarf ellipticals and disks of spiral galaxies (Freeman, 1970).

\item{\bf de Vaucouleurs profile:} It is an r$^{1/4}$ profile. This is a good approximation to the large scale profiles of bright ellipticals and bulge of spirals.

\item{\bf Nuker law:} This is a five parameter model proposed by Lauer et al.~(1995) which is a blend of two power-laws. The Nuker law is used to fit the inner parts of the galactic profiles. 

\end{itemize}

Some of the galaxies are a clear example of a particular profile: e.g. NGC\,5283, follows a perfect de Vaucouleurs law; ESO\,362-G8 matches a Nuker law; and NGC\,4725 shows a Nuker law within the inner 2$^{\prime\prime}$ and a clear exponential in the outer regions. However, less than half of the objects show a correspondence with these profiles. More complex profiles are due to the effects of dust obscuration, occurrence of star-forming regions, or the presence of a bright nucleus, which can dominate the profile up to 2$^{\prime\prime}$. NGC\,5194 is an example of central dust obscuration, while NGC\,5135 shows an irregular profile due to its nuclear starburst. The profiles can show as well the presence of a ring-like structure, as in NGC\,4303 or NGC\,7496. The objects whose nucleus is obscured in the UV images show  a very irregular profile. The classification of the profiles is shown in the last column of Table~\ref{tab:sample_measurements}.

\begin{figure}
\figurenum{9}
\addtocounter{figure}{1}
\plotone{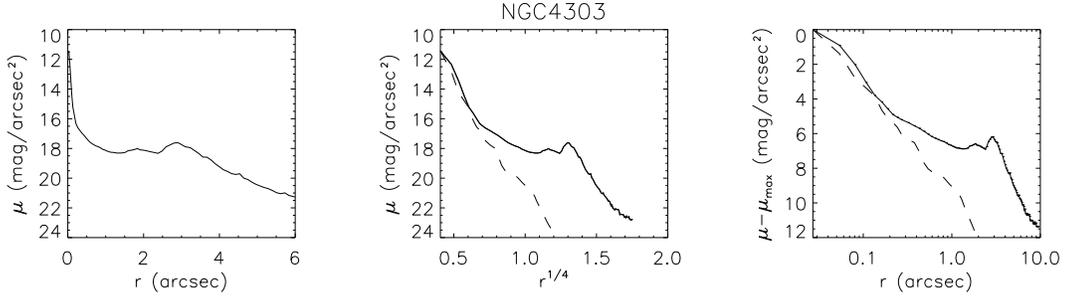}
\caption{Surface brightness profiles are plotted for all the galaxies in the sample. From left to right, the abscissae are scaled linearly, with r$^{1/4}$, and logarithmic, in order to show the type of profile dominating in each galaxy (exponential, de Vaucouleurs or Nuker law). Theoretical PSF profiles are overplotted in dashed line, for comparison with each object. (The whole set of figures is available in the electronic edition of the Journal.)}
\end{figure}


\subsection{Photometry}

The differential surface brightness were obtained by calculating the flux within a very narrow circular annulus, dividing it by the area of the region, and then calculating the magnitude. From the luminosity profiles and the background, we can define the maximum radius (R$_{max}$) as the distance from the centre at which the differential surface brightness equals the background value plus 1$\sigma$.  We consider that this criterion limits the region in which the flux can be calculated with enough S/N, and thus it gives an idea of the extension of the object. We have calculated the magnitude inside apertures of radii R$_{max}$, $1^{\prime\prime}$ and $0.3^{\prime\prime}$. The last two enclose respectively 94\% and 86\% of the total flux for a point-like source. Also, we have calculated the absolute magnitudes inside appertures of projected radii equal to 100\,pc and 300\,pc. All the magnitudes are calculated in the STMAG system, with the formula
\begin{displaymath}
m=-2.5\cdot log(counts/s\cdot PHOTFLAM) -21.1,
\end{displaymath}
where PHOTFLAM is the inverse sensitivity (see Pavlovsky et al.~ 2004, for an explanation of the STMAG system and the calculation of the zero point).

For some objects that are too extended, R$_{max}$ is larger than the distance from the nucleus to the border of the image, and thus it could not be calculated with the standard procedure. We then take an alternative maximum radius that fitted inside the field of view but did not include the borders of the image, where the data do not have enough quality. This is the case of NGC\,3031, NGC\,5941 and NGC\,5005, in which R$_{max}$ becomes a lower limit and thus it affects too, the other measurements of size and magnitude. Note also, that as R$_{max}$ is calculated doing an azimuthal average some emitting features or isolated star forming regions may fall outside the region we are studying. 

 We also calculate the differential surface brightness at $0.3^{\prime\prime}$ and $1^{\prime\prime}$ ($\mu_{0.3^{\prime\prime}}$, $\mu_{1^{\prime\prime}}$), as well as at the half-light radius (radius enclosing half of the total flux, or $\mu_{50}$) , and the radii enclosing 80\%, 50\%, and 20\% of the total flux ($R_{80}$, $R_{50}$, and $R_{20}$). We have obtained that for several objects half or more of the flux is enclosed within a radius of one pixel. For these objects (all the Sy1 and some intermediate type Sy), which exhibit a bright point-like nucleus, we only can set an upper limit for $R_{50}$ and $R_{20}$. All the magnitudes are then corrected for galactic reddening using the extintion coefficients given in Siriani et al.~(2005), which are calculated using the extintion law of Cardelli et al.~(1989). The correction for a particular filter depends on the shape of the continuum. We have used A(F330W)$/$E(B--V)$=$5.054, that is an average of the values given for Sc and elliptical galaxies in Siriani et al.~(2005). The photometry results are given in \mbox{Table~\ref{tab:sample_measurements}}.



\begin{deluxetable}{lccccccccccccccccc}
\rotate
\tabletypesize{\tiny}
\tablewidth{0pc}
\tablecaption{Measurements and results}
\tablehead{
  \colhead{Galaxy}   &
  \colhead{m(0.3$^{\prime\prime}$)} &
  \colhead{m(1$^{\prime\prime}$)}   &
  \colhead{m(R$_{max}$)} &
  \colhead{M(100\,pc)\tablenotemark{a}} &
  \colhead{M(300\,pc)\tablenotemark{a}} &
  \colhead{$\mu (0.3^{\prime\prime})$} &
  \colhead{$\mu (1^{\prime\prime})$} &
  \colhead{$\mu _{50}$} &
   \colhead{R$_{max}$}  &
  \colhead{R$_{80}$}  &
  \colhead{R$_{50}$}  &
 \colhead{R$_{20}$}  &  
  \colhead{Profile}
  \\
  \colhead{Name}     &
  \colhead{}    &
  \colhead{}     &
  \colhead{} &
  \colhead{} &
  \colhead{} &
  \colhead{} & 
  \colhead{} & 
  \colhead{} & 
  \colhead{pc\,($^{\prime\prime}$)} &
  \colhead{pc\,($^{\prime\prime}$)} &
  \colhead{pc\,($^{\prime\prime}$)} &
  \colhead{pc\,($^{\prime\prime}$)} &
  \colhead{}   
\\
  \colhead{(1)}     &
  \colhead{(2)}     &
  \colhead{(3)}    &
  \colhead{(4)}     &
  \colhead{(5)} &
  \colhead{(6)} &
  \colhead{(7)} &
  \colhead{(8)} &
  \colhead{(9)} &
  \colhead{(10)} &
  \colhead{(11)} &
  \colhead{(12)} &
  \colhead{(13)} &        
  \colhead{(14)}         
  }
\startdata
CGCG\,164-019                  & 17.66 & 17.05 &  16.47 & -17.50  & -17.96 & 17.98  & 19.48 &  19.39 & 1586 (2.74)    & 1013 (1.75)  & (439) 0.76    &  52 (0.09)   & v  \\
Circinus                       & 14.12 & 11.88 &   9.71 &   --    &   --   & 12.93  & 13.29 &  13.43 &   62 (3.23)    &   53 (2.77)  &  (40) 2.12    &  24 (1.27)   & 0  \\
ESO\,103-G35                   & 19.79 & 18.06 &  17.01 & -14.20  & -15.79 & 18.93  & 19.69 &  19.92 &  578 (2.25)    &  475 (1.85)  & (315) 1.23    & 157 (0.61)   & n  \\
ESO\,137-G34                   & 19.83 & 17.21 &  16.11 & -14.55  & -16.47 & 18.36  & 18.77 &  18.82 &  354 (1.99)    &  299 (1.68)  & (221) 1.24    & 123 (0.69)   & 0  \\
ESO\,138-G1                    &  16.4 & 15.43 &  15.09 & -17.00  & -17.62 & 15.56  & 18.11 &  16.92 &  390 (2.20)    &  216 (1.22)  &  (95) 0.54    &  37 (0.21)   & np  \\
ESO\,362-G8                    & 17.66 & 15.85 &  14.34 & -16.50  & -18.13 & 16.51  & 17.70 &  18.97 & 1906 (6.17)    & 1307 (4.23)  & (659) 2.13    & 256 (0.83)   & n  \\
fairall49                      & 18.03 & 16.75 &  15.98 & -16.38  & -17.49 & 17.66  & 18.73 &  18.74 &  930 (2.37)    &  604 (1.54)  & (398) 1.01    & 165 (0.42)   & er  \\
IC\,2560                       & 16.91 & 16.36 &  16.02 & -16.33  & -16.81 & 16.75  & 19.28 &  17.32 &  409 (2.16)    &  231 (1.22)  &  (72) 0.38    &  21 (0.11)   & 0  \\
IC\,4870                       & 15.54 & 15.34 &  14.48 & -15.23  &    --  &  17.2  & 18.84 &  19.05 &  283 (4.96)    &  192 (3.37)  &  (78) 1.37    & $<$3 (0.05)  & enp  \\
IC\,5063                       & 19.23 &  17.9 &  15.81 & -14.47  & -15.85 & 18.74  & 19.66 &  19.92 &  876 (3.98)    &  722 (3.28)  & (507) 2.30    & 277 (1.26)   & e  \\
Mrk\,6                         & 14.85 & 14.67 &  14.57 & -19.53  & -19.68 & 16.31  & 18.79 &  11.32 &  732 (2.00)    &  139 (0.38)  & $<$(24) 0.07  & $<$11 (0.03) & np  \\
Mrk\,40                        & 16.99 & 16.72 &  16.58 & -17.61  & -17.83 & 18.22  & 20.30 &  14.52 &  691 (1.69)    &  274 (0.67)  &  (38) 0.09    & $<$12 (0.03) & np  \\
Mrk\,42                        & 15.92 & 15.67 &  15.55 & -19.01  & -19.14 &  17.8  & 18.57 &  12.35 &  796 (1.67)    &  348 (0.73)  & $<$(32) 0.07  & $<$14 (0.03) & 0pr  \\
Mrk\,231                       & 14.54 & 14.43 &  14.18 & -21.44  & -21.61 & 16.57  & 18.88 &  11.02 & 3749 (4.59)    &  931 (1.14)  & $<$(55) 0.07  & $<$25 (0.03) & vp  \\
Mrk\,334                       & 17.05 & 16.31 &  15.32 & -17.59  & -18.19 & 17.61  & 18.96 &  19.91 & 1760 (4.14)    & 1309 (3.08)  & (700) 1.65    & 119 (0.28)   & 0  \\
Mrk\,461                       & 15.88 & 15.58 &  15.13 & -18.24  & -18.51 &  16.5  & 19.16 &  16.47 & 1133 (4.32)    &  590 (2.25)  &  (77) 0.30    &  21 (0.08)   & 0r  \\
Mrk\,471                       & 19.58 & 18.49 &  16.28 & -15.76  & -16.39 & 19.57  & 20.07 &  20.75 & 2758 (4.16)    & 2400 (3.62)  &(1741) 2.63    & 922 (1.39)   & 0p  \\
Mrk\,477                       &  16.5 & 16.03 &  15.44 & -19.42  & -19.84 & 16.61  & 19.13 &  18.32 & 1080 (3.44)    &  703 (2.24)  & (203) 0.64    &  41 (0.13)   & 0  \\
Mrk\,493                       & 14.71 &  14.5 &  14.43 & -20.50  & -20.66 & 16.58  & 18.55 &  11.49 & 1153 (1.90)    &  261 (0.43)  & $<$(44) 0.07  & $<$18 (0.03) & 0pr  \\
Mrk\,516                       & 18.54 & 17.83 &  17.14 & -16.07  & -17.11 &  17.7  & 20.23 &  20.17 & 1185 (2.15)    &  865 (1.57)  & (509) 0.92    & 127 (0.23)   & 0  \\
Mrk\,915                       & 15.11 & 14.94 &  14.79 & -19.71  & -19.85 & 16.89  & 18.81 &  11.57 & 1137 (2.43)    &  262 (0.56)  & $<$(30) 0.06  & $<$14 (0.03) & 0p  \\
Mrk\,1210                      & 16.54 & 16.11 &   15.4 & -16.76  & -17.24 & 16.94  & 19.20 &  19.04 & 2866 (3.92)    & 2047 (2.80)  & (667) 0.91    &  73 (0.10)   & 0r  \\
NGC\,449                       & 17.45 & 16.95 &  16.51 & -16.63  & -17.06 & 17.31  & 20.02 &  18.33 &  699 (2.26)    &  467 (1.51)  & (133) 0.43    &  34 (0.11)   & 0  \\
NGC\,1144                      & 19.81 & 18.22 &  16.26 & -14.87  & -16.29 & 18.91  & 20.07 &  21.67 & 2851 (5.10)    & 2622 (4.69)  &(2029) 3.63    & 688 (1.23)   & 0  \\
NGC\,1320                      & 17.64 & 16.92 &  15.39 & -15.48  & -16.30 & 17.61  & 19.24 &  20.14 &  893 (5.19)    &  702 (4.08)  & (419) 2.43    & 127 (0.74)   & v  \\
NGC\,1672                      & 18.78 & 16.42 &  12.39 & -15.08  & -17.28 & 17.41  & 17.92 &  17.69 &  875 (10.18)   &  587 (6.83)  & (423) 4.92    & 280 (3.25)   & 0  \\
NGC\,2639                      & 19.32 & 17.47 &  15.04 & -14.70  & -16.21 & 18.06  & 19.28 &  20.19 & 1326 (6.14)    & 1084 (5.02)  & (705) 3.26    & 352 (1.63)   & n  \\
NGC\,3031\tablenotemark{b,c}   & 15.79 & 15.07 &  17.48 & -14.90  &    --  & 16.34  & 17.23 &  18.54 & $>$185 (10.85) &  145 (8.52)  &  (93) 5.45    &  43 (2.52)   & ep  \\
NGC\,3081                      & 17.76 & 16.58 &  14.71 & -15.33  & -16.54 & 17.41  & 18.50 &  20.34 &   979 (6.36)   &  788 (5.12)  & (489) 3.17    & 174 (1.13)   & 0  \\
NGC\,3227                      & 14.46 & 14.27 &  14.05 & -16.74  &    --  & 15.85  & 18.16 &  12.11 &   283 (3.79)   &   65 (0.87)  &   (8) 0.10    & $<$2 (0.03)  & 0p  \\
NGC\,3362                      & 19.22 & 18.35 &  16.52 & -15.64  & -16.45 & 19.01  & 20.77 &  21.79 &  2615 (4.88)   & 2385 (4.45)  &(1781) 3.32    & 600 (1.12)   & nr  \\
NGC\,3393                      & 18.67 & 16.75 &  14.95 & -15.35  & -17.08 & 17.46  & 18.30 &  18.66 &  1135 (4.69)   &  784 (3.24)  & (466) 1.93    & 249 (1.03)   & n  \\
NGC\,3486                      & 18.05 & 17.14 &   15.7 & -13.25  &   --   & 17.95  & 19.31 &  19.95 &   148 (4.12)   &  116 (3.23)  &  (72) 1.99    &  26 (0.71)   & e  \\
NGC\,3516                      & 13.67 & 13.51 &  13.11 & -19.15  & -19.35 & 15.54  & 17.39 &  12.04 &  1014 (5.93)   &  347 (2.03)  &  (20) 0.11    & $<$5 (0.03)  & np  \\
NGC\,3786                      & 17.07 & 16.37 &  15.96 & -16.03  & -16.70 & 17.32  & 18.60 &  18.33 &   381 (2.20)   &  220 (1.27)  & (104) 0.60    &  14 (0.08)   & 0  \\
NGC\,3982                      & 18.44 & 17.44 &  15.38 & -13.94  & -15.14 & 18.75  & 19.35 &  20.61 &   524 (6.39)   &  436 (5.32)  & (263) 3.21    & 107 (1.30)   & 0  \\
NGC\,4253                      & 15.61 &  15.3 &  14.75 & -17.98  & -18.27 & 16.41  & 18.69 &  17.31 &  1199 (4.89)   &  684 (2.79)  & (115) 0.47    &  15 (0.06)   & np  \\
NGC\,4258                      & 17.86 & 15.92 &   13.2 & -15.02  &   --   & 16.78  & 17.49 &  18.38 &   205 (6.20)   &  163 (4.93)  & (112) 3.39    &  59 (1.79)   & en  \\
NGC\,4303                      & 16.19 & 15.56 &  13.54 & -16.03  & -17.50 & 16.61  & 18.02 &  17.60 &   558 (5.52)   &  361 (3.57)  & (277) 2.74    & 141 (1.40)   & 0pr  \\
NGC\,4395                      & 16.77 & 16.57 &   16.5 &   --    &   --   & 17.43  & 20.78 &  14.40 &    22 (1.43)   &    5 (0.32)  &   (2) 0.10    & $<$1 (0.04)  & 0  \\
NGC\,4565                      & 19.91 & 18.27 &  17.16 & -13.18  &   --   & 19.14  & 19.86 &  20.00 &   181 (2.26)   &  144 (1.80)  & (100) 1.25    &  52 (0.65)   & 0  \\
NGC\,4593                      & 13.33 & 13.22 &  13.03 & -19.51  & -19.63 & 15.35  & 17.64 &   9.64 &   813 (4.67)   &   87 (0.50)  & $<$(11) 0.06  & $<$5 (0.03)  & 0p  \\
NGC\,4725                      & 18.26 & 16.58 &     14 & -14.80  & -16.23 & 17.31  & 18.29 &  19.64 &   651 (8.35)   &  496 (6.36)  & (313) 4.02    & 144 (1.84)   & e  \\
NGC\,4939                      & 18.98 & 17.08 &  16.27 & -14.99  & -16.33 & 18.06  & 19.21 &  19.44 &   553 (2.75)   &  410 (2.04)  & (215) 1.07    & 103 (0.51)   & v  \\
NGC\,4941                      & 18.02 & 16.93 &  15.58 & -14.26  &   --   & 17.26  & 18.95 &  19.60 &   279 (3.87)   &  211 (2.93)  & (128) 1.78    &  48 (0.67)   & e  \\
NGC\,5005\tablenotemark{b}     & 18.15 & 16.38 &  22.32 & -15.22  & -16.54 & 17.06  & 18.11 &  20.00 &$>$1252 (12.15) & 1049 (10.18) & (730) 7.09    & 333 (3.23)   & ev  \\
NGC\,5033                      &    16 & 15.64 &  13.65 & -15.77  & -16.45 & 17.04  & 18.58 &  20.14 & 1060 (11.64)   &  844 (9.27)  & (520) 5.72    & 148 (1.63)   & np  \\
NGC\,5135                      & 16.49 &  15.4 &  13.59 & -17.39  & -18.52 & 16.08  & 17.47 &  17.44 & 1721 (6.47)    & 1064 (4.00)  & (470) 1.77    & 277 (1.04)   & 0  \\
NGC\,5194\tablenotemark{b,c}   & 17.73 & 15.34 &  13.44 & -14.74  & -15.83 & 16.09  & 17.56 &  19.73 & $>$340 (9.18)  &  277 (7.48)  & (173) 4.67    &  38 (1.03)   & 0  \\
NGC\,5256                      & 17.07 & 16.54 &  16.11 & -17.81  & -18.51 & 16.86  & 19.79 &  17.42 &  1468 (2.72)   &  896 (1.66)  & (223) 0.41    &  70 (0.13)   & 0  \\
NGC\,5273                      & 15.29 & 15.02 &  14.65 & -16.32  & -16.71 & 16.83  & 18.41 &  17.19 &   388 (3.76)   &  162 (1.57)  &  (43) 0.41    & $<$3 (0.03)  & np  \\
NGC\,5283                      & 18.24 & 16.92 &     16 & -15.45  & -16.57 & 17.34  & 18.96 &  19.13 &   582 (2.88)   &  418 (2.07)  & (237) 1.18    &  91 (0.45)   & v  \\
NGC\,5347                      & 18.13 & 17.15 &  16.88 & -15.00  &   --   & 17.79  & 19.80 &  18.51 &   267 (1.76)   &  159 (1.05)  &  (83) 0.55    &  24 (0.16)   & 0  \\
NGC\,5548                      & 14.31 & 14.17 &  13.98 & -19.87  & -20.00 & 15.93  & 18.35 &  10.84 &  1213 (3.65)   &  216 (0.65)  & $<$(22) 0.07  & $<$10 (0.03) & 0p  \\
NGC\,5674                      & 18.05 & 16.97 &  16.61 & -16.65  & -17.65 & 17.63  & 19.67 &  18.32 &   966 (2.00)   &  599 (1.24)  & (292) 0.60    &  97 (0.20)   & 0  \\
NGC\,5695                      & 19.11 &  17.6 &   16.6 & -14.90  & -16.26 & 18.18  & 19.48 &  19.94 &   767 (2.81)   &  584 (2.14)  & (344) 1.26    & 147 (0.54)   & v  \\
NGC\,5728                      &  21.1 & 17.52 &  14.42 & -13.67  & -16.35 & 19.48  & 18.69 &  19.22 &  1148 (6.38)   &  846 (4.70)  & (656) 3.64    & 369 (2.05)   & 0r  \\
NGC\,5940                      & 15.43 & 15.31 &  15.26 & -20.12  & -20.28 & 17.36  & 19.74 &  11.63 &  1042 (1.59)   &  118 (0.18)  & $<$(40) 0.06  & $<$20 (0.03) & 0p  \\
NGC\,6300                      &  21.1 & 18.62 &  17.62 & -12.81  &   --   &  19.6  & 19.99 &  20.00 &   123 (1.77)   &  106 (1.53)  &  (78) 1.13    &  46 (0.67)   & 0  \\
NGC\,6814                      & 13.25 & 13.13 &  13.02 & -18.27  &   --   & 15.26  & 17.47 &   9.93 &   279 (2.79)   &   25 (0.25)  & $<$(7) 0.07   & $<$3 (0.03)  & 0p \\
NGC\,6951                      & 18.23 & 16.9  &  13.74 & -14.60  & -16.55 & 17.25  & 18.73 &  18.64 &   474 (5.15)   &  417 (4.53)  & (344) 3.74    & 222 (2.41)   & 0r  \\
NGC\,7130                      & 15.98 & 15.17 &  13.47 & -18.13  & -18.87 & 15.53  & 18.64 &  20.04 &  3333 (10.65)  & 2601 (8.31)  &(1908) 6.09    & 266 (0.85)   & 0  \\
NGC\,7212                      & 18.03 & 17.19 &  16.43 & -16.47  & -17.59 & 17.2   & 19.53 &  19.54 &  1310 (2.54)   &  980 (1.90)  & (518) 1.00    & 134 (0.26)   & 0  \\
NGC\,7319                      & 20.51 & 18.91 &  18.91 & -14.05  & -15.41 & 20.15  & 20.86 &  19.84 &   441 (1.01)   &  349 (0.80)  & (247) 0.57    & 109 (0.25)   & 0  \\
NGC\,7469                      & 13.22 & 13.08 &  12.65 & -20.85  & -20.98 & 15.14  & 16.85 &  11.83 &  1376 (4.35)   &  506 (1.60)  &  (39) 0.12    & $<$9 (0.03)  & 0pr  \\
NGC\,7479                      & 19.56 &    18 &  16.89 & -14.01  & -15.34 & 18.49  & 19.91 &  20.05 &   367 (2.38)   &  310 (2.01)  & (205) 1.33    &  89 (0.58)   & 0  \\
NGC\,7496                      & 16.48 & 15.43 &  14.35 & -16.23  & -17.18 & 15.75  & 17.55 &  17.40 &   538 (5.02)   &  272 (2.54)  & (143) 1.33    &  45 (0.42)   & 0  \\
NGC\,7674                      & 16.47 & 15.89 &  15.69 & -18.52  & -19.16 &  16.4  & 18.89 &  16.48 &  1082 (1.93)   &  510 (0.91)  & (177) 0.32    &  50 (0.09)   & n  \\
NGC\,7743                      & 17.22 & 16.01 &  15.04 & -15.71  & -16.48 & 16.38  & 18.41 &  18.90 &   463 (4.18)   &  316 (2.85)  & (152) 1.37    &  48 (0.43)   & n  \\
UGC\,1214                      & 17.44 & 16.51 &  15.41 & -16.75  & -17.60 & 16.85  & 18.87 &  19.43 &  1333 (3.99)   &  945 (2.83)  & (522) 1.56    & 137 (0.41)   & n  \\
UGC\,1395                      & 18.52 & 17.35 &  17.26 & -15.69  & -16.79 &  18.4  & 20.06 &  18.46 &   381 (1.23)   &  270 (0.80)  & (177) 0.52    &  40 (0.12)   & 0  \\
UGC\,2456                      & 16.53 & 15.85 &  15.36 & -17.11  & -17.76 & 15.84  & 18.31 &  17.75 &   598 (2.57)   &  333 (1.43)  & (143) 0.61    &  44 (0.19)   & 0  \\
UGC\,6100                      & 18.84 & 17.84 &  16.63 & -15.94  & -16.99 & 18.28  & 20.11 &  21.03 &  2219 (3.88)   & 1830 (3.20)  &(1022) 1.79    & 297 (0.52)   & n  \\
UGC\,12138                     & 14.96 & 14.73 &  14.62 & -19.98  & -20.19 & 16.38  & 18.88 &  11.64 &  1054 (2.18)   &  232 (0.48)  & $<$(34) 0.07  & $<$15 (0.03) & np  \\
UM\,625                        & 17.44 & 16.35 &  16.15 & -17.39  & -18.29 & 17.53  & 19.39 &  17.37 &   887 (1.83)   &  436 (0.90)  & (252) 0.52    &  53 (0.11)   & 0  \\
\enddata
\label{tab:sample_measurements}
\tablecomments{Col.\ (1): Galaxy name; Col.\ (2): Magnitude within 0.3$^{\prime\prime}$ radius.
 Col.\ (3): Magnitude within 1$^{\prime\prime}$ radius. Col.\ (4):  Magnitude within the maximum radius. Col.\ (5): Absolute magnitude within a projected radius of 100\,pc. Col.\ (6): Absolute magnitude within 300\,pc.
Col.\ (7) \& (8): Differential surface brightness at 0.3$^{\prime\prime}$ and 1$^{\prime\prime}$ respectively. Col.\ (9): Differential surface brightness at the half-light radius. Col.\ (10): Computed maximum radius in pc and arcsec in brackets.  Col.\ (11), (12) \& (13): Radius enclosing 80\%, 50\% and 20\% of the flux within R$_{max}$, in pc (same in arcsec in brackets). Col.\ (14): Classification of the profiles from Figs. 9.1-9.75. The type of profile is coded with the letters: `e' for an exponential profile, `v' for a de Vaucouleurs law, `n' when a Nuker law is seen, and `0' if the profile does not fall in any of the former categories. A letter `p' is added when there is a point-like nucleus present, and an `r' if there is a ring visible in the image. \\ All the magnitudes are calculated in the STMAG system and corrected for galactic extinction.}  
\tablenotetext{a}{In the cases in which R$_{max}$ is less than 100\,pc, the absolute magnitude (M) within 300\,pc or 100\,pc is not calculated. If R$_{max}$ is less than 300\,pc but greater than 100\,pc, then M within 300\,pc is computed, but not M within 100\,pc.}
\tablenotetext{b}{In these cases R$_{max}$ is limited by the border of the field of view and not by the integration.}
\tablenotetext{c}{Occulting finger of the HRC limits the radius for which the asymmetry parameter (see text) is computed (smaller than R$_{max}$).}
\end{deluxetable}

\begin{figure}
\epsscale{1}
\plotone{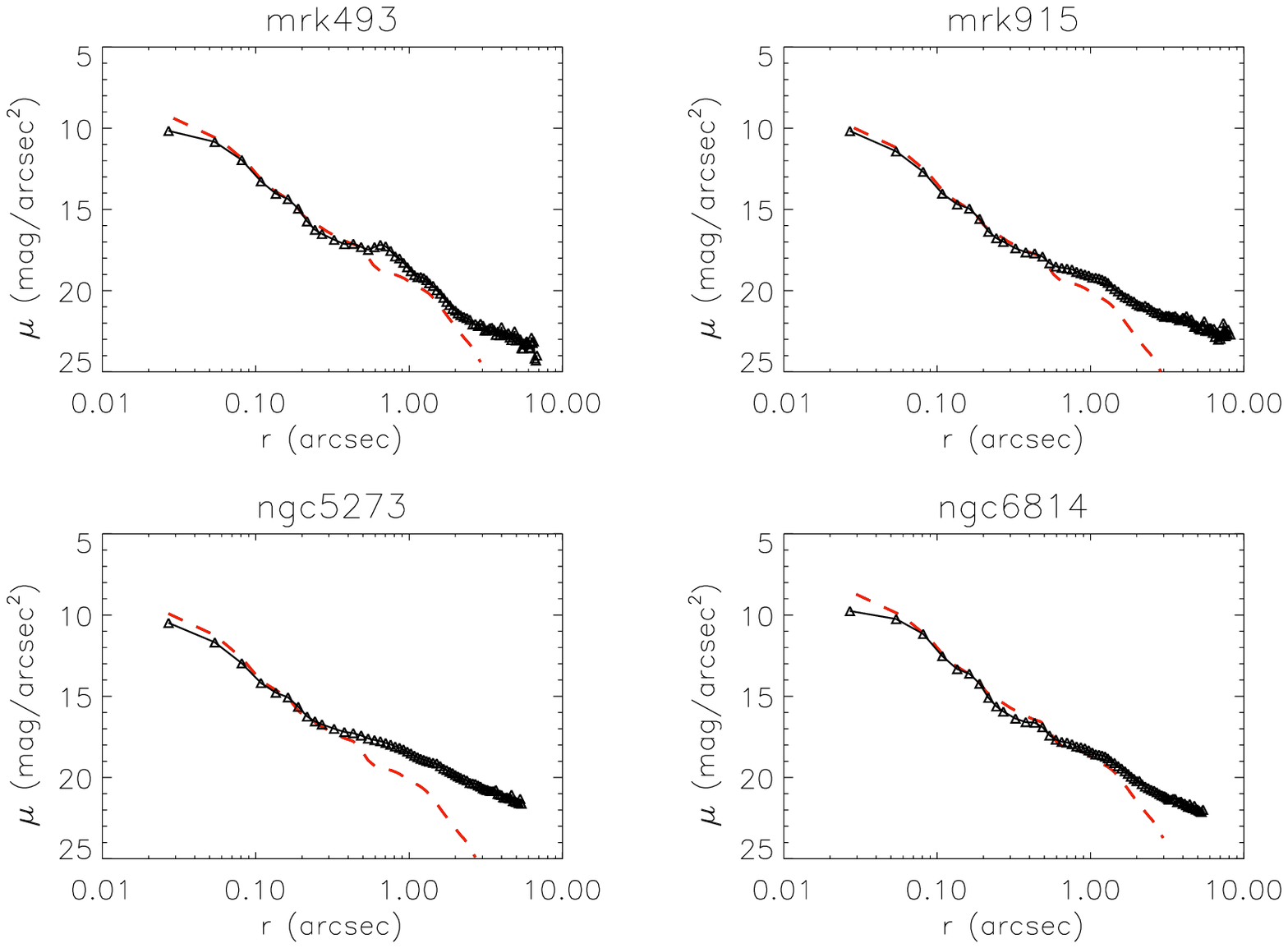}
\caption{PSF fit to the nucleus for the saturated objects with two images. The fitted PSF is the red dashed line, that is overplotted to the surface brightness curve (black line and triangles), for each object. The most affected nucleus is NGC\,6814, while Mrk\,915 has barely lost flux in the saturated image.}
\label{fig:f10}
\end{figure}

As explained above, saturation might affect the flux in some nuclei. This effect can be corrected straightforwardly for the objects overcoming the saturation threshold which have also short exposure images. First we checked that from a radius of 0.1$^{\prime\prime}$ outwards, the surface brightness profiles calculated from both images were coincident. Then the inner 8 pixels of the saturated and high S/N image were replaced by the data from the unsaturated image. The final analysis was carried out on this corrected profile. For the saturated objects with just one image we performed a $\chi^2$ fit of a Tinytim-generated PSF to the wings of the nuclear PSF. The fit was done in the range of 5--12 pixels (0.13$^{\prime\prime}$--0.32$^{\prime\prime}$), range in which the pixels are not severely affected by the saturation and the S/N is still high. We considered possible focus changes by allowing the PSF to be broadened up to a 10\%, choosing the best fit from the whole set of different broadenings. The inner 0.2$^{\prime\prime}$ of the galaxy was replaced by the fitted PSF. We checked this method with those core-saturated galaxies which have additional non-saturated exposures, obtaining a good agreement with the fluxes calculated from the combined profiles. Fig.~\ref{fig:f10} show the fitted PSF for these four nuclei. The resulting corrections calculated range from 0.06 mag (Mrk\,915) to 0.36 mag (NGC\,5273). The nucleus of NGC\,5940 does not reach the saturation threshold, but it is close to it. It has only a long exposure image, in which it does not seem to be affected by saturation. The correction that we expect for this object should be smaller than that for Mrk\,493 or Mrk\, 915, which have brighter nuclei.

\begin{figure}
\plotone{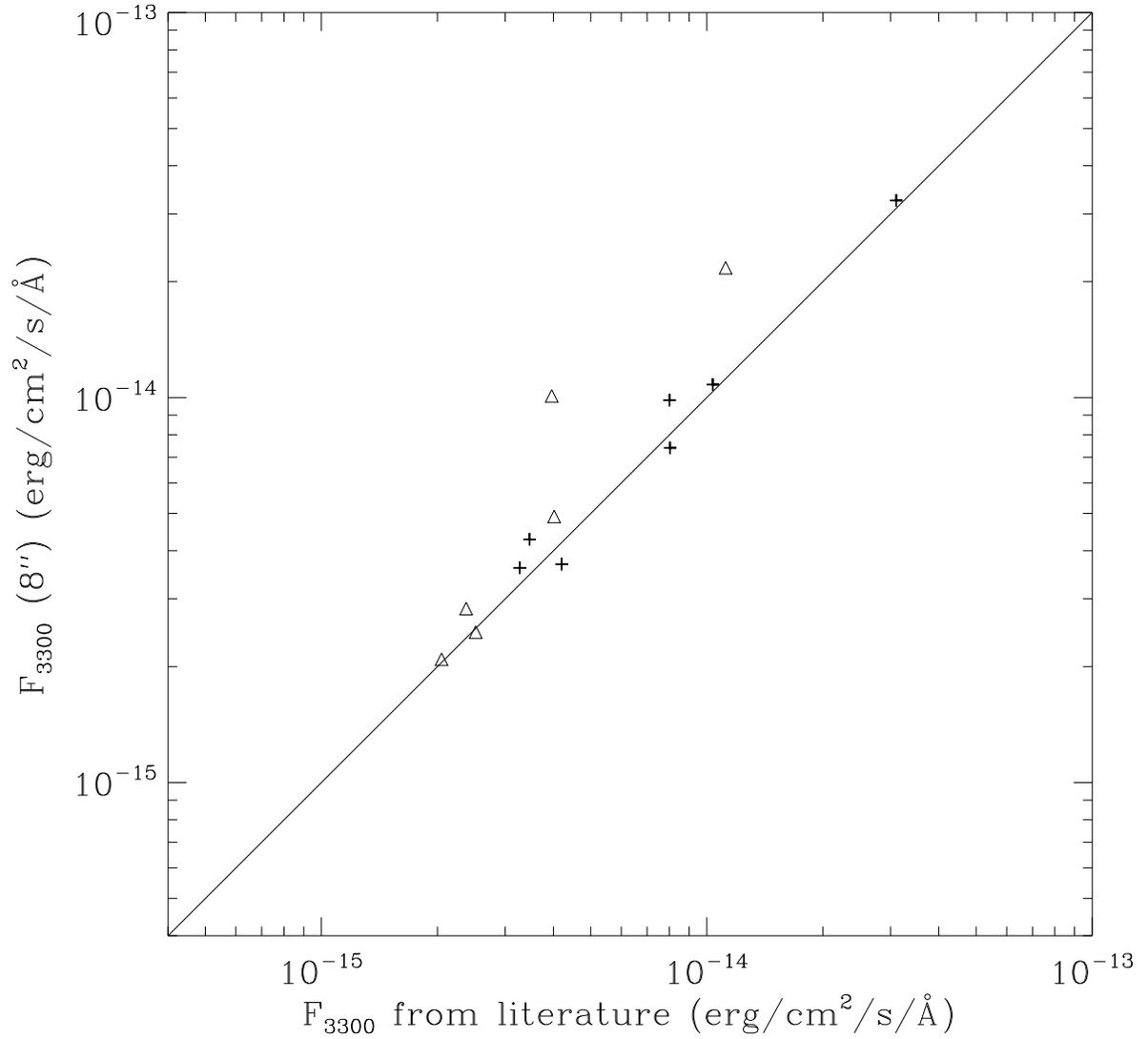}
\caption{Comparison of the fluxes that we have measured with data from the literature. Triangles represent fluxes from Kinney et al.~(1993), while crosses represent fluxes from Storchi-Bergmann et al.~(1995). The two triangles that fall over the unity line are NGC\,4258 and NGC\,5005 (see discussion in text).}
\label{fig:f11}
\end{figure}

The magnitudes given in Table~\ref{tab:sample_measurements} depend on the determination of R$_{max}$, so one has to take this into account when using these fluxes. In order to illustrate this we have compared the total fluxes of some of our galaxies with the fluxes presented in Storchi-Bergmann et al.~(1995) and Kinney et al.~(1993). The compared subsamples are: NGC\,3982, NGC\,4258, NGC\,5005, NGC\,5256, NGC\,5674 and Mrk\,477, that have been studied by Kinney et al.~(1993) and have published fluxes at $\sim$2700\AA\, and spectral slopes ($\beta$); and the subsample NGC\,3081, NGC\,3393, NGC\,5135, NGC\,5728, NGC\,1672, NGC\,7130 and NGC\,7496, that have been studied as well by Storchi-Bergmann et al.~(1995), giving fluxes at 2900 and 3500\AA. \mbox{Fig.~\ref{fig:f11}} shows the comparison between their measurements and ours. A good general agreement is found. When comparing the fluxes several issues have to be taken into account, the different instrumental set-up being the most determining. In order to have a good S/N we have measured inside a radius R$_{max}$, while they used the aperture of the IUE slit, that is 10$^{\prime\prime} \times$20$^{\prime\prime}$. This is equivalent in surface to a circular aperture of 8$^{\prime\prime}$, although the flux depends on the light distribution of the object and the orientation of the slit. Thus, when we use apertures of 8$^{\prime\prime}$ the agreement is very good, except for NGC\,4258 and NGC\,5005, for which we measure a flux 1.9 and 2.5 times higher respectively. In these cases the isophotes are clearly elongated, so the flux is expected to vary significantly with the slit orientation. Moreover, because the calculation of R$_{max}$ implies an azimuthal average then some bright features can lie out of this region. One example of this is NGC\,5674, that has an external ring-like structure which is outside of R$_{max}$, but fits in a 8$^{\prime\prime}$ aperture. The fluxes at 3300\AA\, have been interpolated between the fluxes at 2900\AA\, and 3500\AA\, given by Storchi-Bergmann et al.~(1995), or corrected from the fluxes at 2700\AA\, given by Kinney et al.~(1993) using the value of $\beta$ that they calculate. In Table~\ref{tab:compara_kinney} we list the calculated photometric values for these objects.

\begin{deluxetable}{lcccc}
\tabletypesize{\small}
\tablewidth{0pc}
\tablecaption{Comparison of flux measurements.}
\tablehead{
  \colhead{\textbf{Name}}   &
  \colhead{\textbf{R$_{max}$}}   &
  \colhead{\textbf{F$_{3300}$ (R$_{max}$)}}  &
  \colhead{\textbf{F$_{3300}$ (8$^{\prime\prime}$)}}  &
  \colhead{\textbf{F$_{3300}$ (10$\times$20$^{\prime\prime}$)}}   
\\
  \colhead{\textbf{ }}   &
  \colhead{\textbf{($^{\prime\prime}$)}}   &
  \colhead{\textbf{ }}  &
  \colhead{\textbf{ }}  &
  \colhead{\textbf{ }}   
  }
\startdata
Mrk477\tablenotemark{a}   & 3.4  &  2.3   &  2.8  & 2.4  \\
NGC1672  & 10.2 & 36.2   & 32.5  & 31.0   \\
NGC3081  & 6.4  &  3.7  &  4.3  & 3.5 \\
NGC3393  & 4.7   &  2.7  &  3.6  & 3.3 \\
NGC3982\tablenotemark{a}  & 6.4   &  3.1  &  4.9  & 4.0  \\
NGC4258\tablenotemark{a}  & 6.2   &  13.0    & 21.7  & 11.2 \\
NGC5005\tablenotemark{a}  & 12.2 &  17.0    & 10.1  & 4.0  \\
NGC5135  & 6.5   & 10.0  & 10.8  & 10.4\\
NGC5256\tablenotemark{a}  & 2.7  &  1.2   &  2.5  & 2.5  \\
NGC5674\tablenotemark{a}  & 2.0     &  0.7   &  2.1  & 2.1  \\
NGC5728  & 6.4  &  3.5   &  3.7  & 4.2  \\
NGC7130  & 10.7 &  13.0    &  9.8  & 8.0    \\
NGC7496  & 5.0     &  6.3   &  7.4  & 8.0 
\enddata
\tablecomments{This table shows the comparison between our measurements and the values published by Kinney et al.~(1993) and Storchi-Bergmann et al.~(1995).   \\
Col.\ (1): galaxy name; Col.\ (2): maximum radius; Col.\ (3): UV flux measured at maximum radius; Col.\ (4): UV flux measured at 8$^{\prime\prime}$ radius; Col.\ (5): UV flux from literature at 3300\AA. Units of Col.\ (3) -- (5) are 10$^{-15}$ erg/s/cm$^2$/\AA.}
\tablenotetext{a}{Fluxes in Col. (5) have been calculated using the spectral slope given in Kinney et al.~(1993) for these galaxies.}
\label{tab:compara_kinney}
\end{deluxetable}

\begin{figure}
\plotone{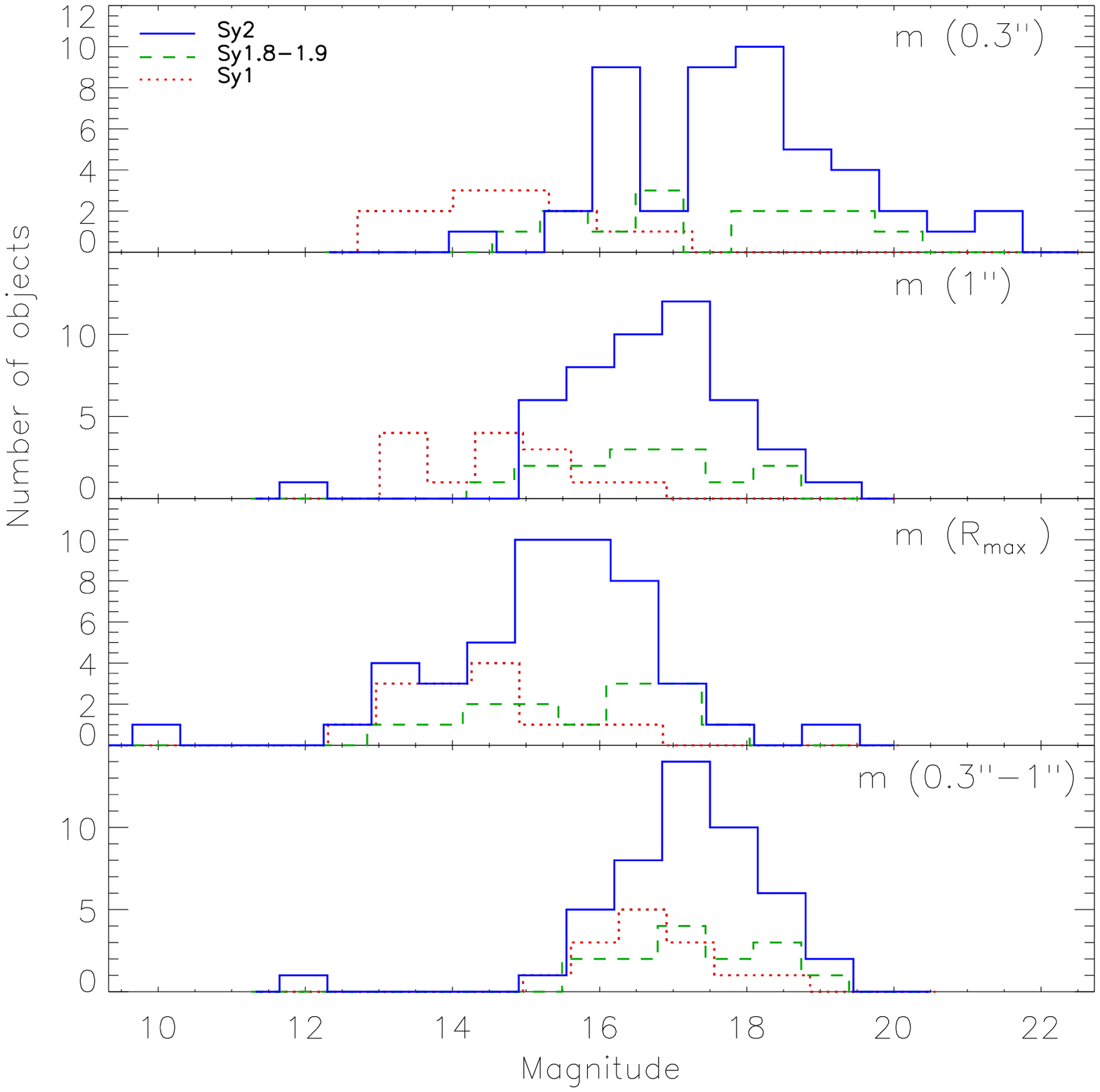}
\caption{Comparative histogram of the magnitudes measured within different radii. Sy2 are plotted in blue full line; Sy1 in red dotted line; and Sy1.8-1.9 in green dashed line. Sy1 nuclei tend to be brighter than the others, although this trend is softened when larger radii are considered. The lower panel shows the magnitude between 0.3 and 1$^{\prime\prime}$ apertures.}
\label{fig:f12}
\end{figure}

\begin{figure}
\plotone{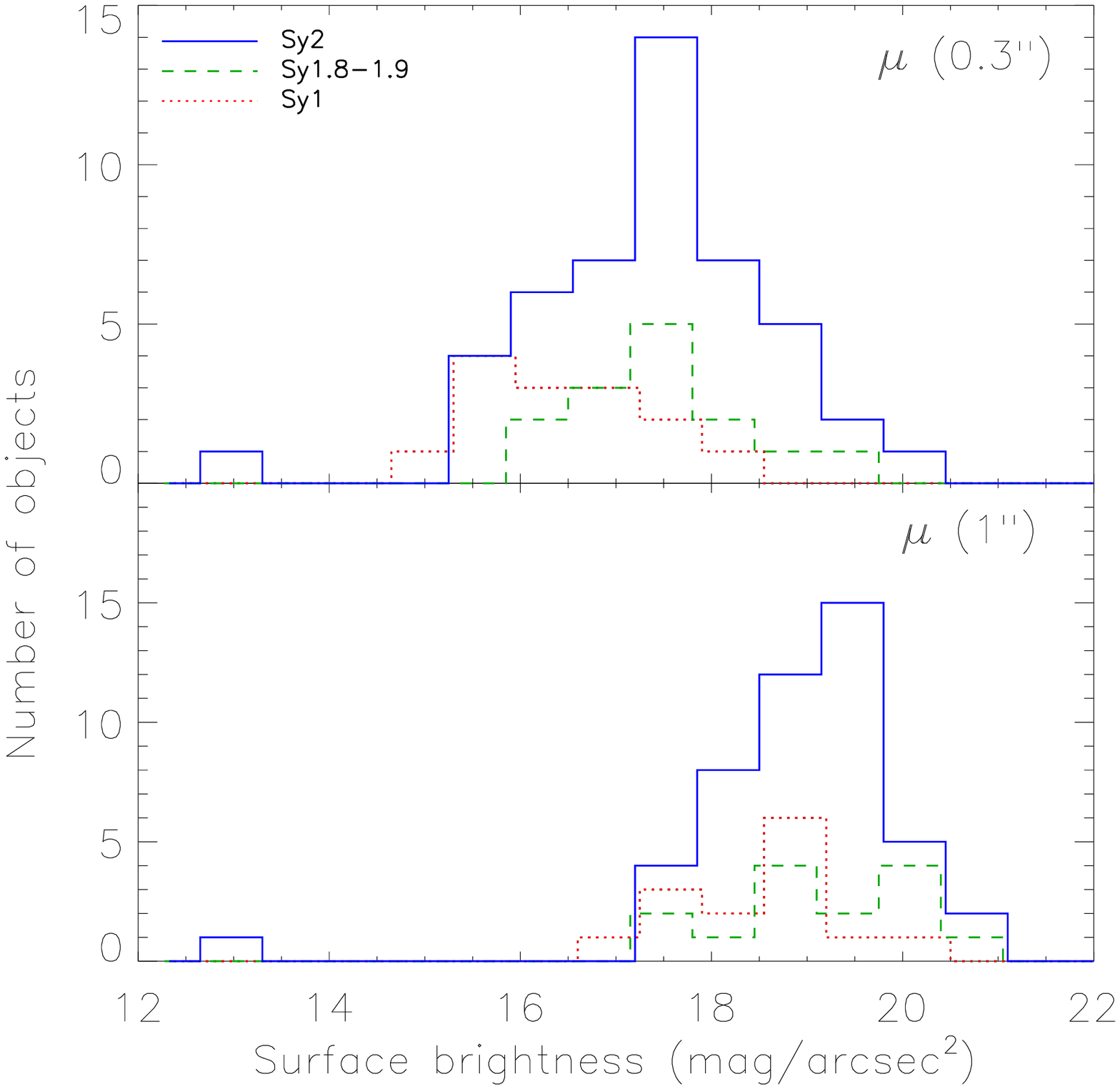}
\caption{Comparative histogram of the values of the surface brightness  measured at two different radii. At 0.3$^{\prime\prime}$ Sy1 nuclei are brighter than those from other types. At 1$^{\prime\prime}$, however, there is no difference in the distribution of $\mu$. Colors and type of line are as in previous figure.}
\label{fig:f13}
\end{figure}

Fig.~\ref{fig:f12} presents comparative histograms of the values of the magnitudes measured within 0.3$^{\prime\prime}$, 1$^{\prime\prime}$, R$_{max}$, and the magnitude in a circular ring between 0.3$^{\prime\prime}$ and 1$^{\prime\prime}$. It is shown that Sy1 nuclei are brighter at small radii. However this trend is not observed when the contribution of the inner 0.3$^{\prime\prime}$ is subtracted, indicating that the light in Sy1 is dominated by the compact nucleus, and the difference of the subsamples in terms of magnitude is not large. The bright outlier object in the plots is Circinus galaxy, that due to its low galactic latitude has a large extinction correction that makes it even brighter than the Messier objects of the sample. In Fig.~\ref{fig:f13} we plot the surface brightness $\mu$ at 0.3$^{\prime\prime}$ and 1$^{\prime\prime}$. The calculation of the surface brightness at 1$^{\prime\prime}$ is practically unaffected by the Sy1 nuclei, what causes the dissimilarity between the panels of Fig.~\ref{fig:f13}. This indicates no significant difference among the host of Sy1 and Sy2, in terms of surface brightness. The difference would be due just to the presence of the nuclear source in Sy1.

\subsection{Compactness and Asymmetry}

The morphology of extended objects can be quantified with the concentration or compactness (C) and asymmetry (A) parameters. Due to the irregular distribution of light at $\lambda$3300, these parameters give a better description of the morphology than the classical bulge-disk decomposition. They also reflect the contribution of clumpy structure, such as star clusters and star-forming regions.

\begin{figure}
\plotone{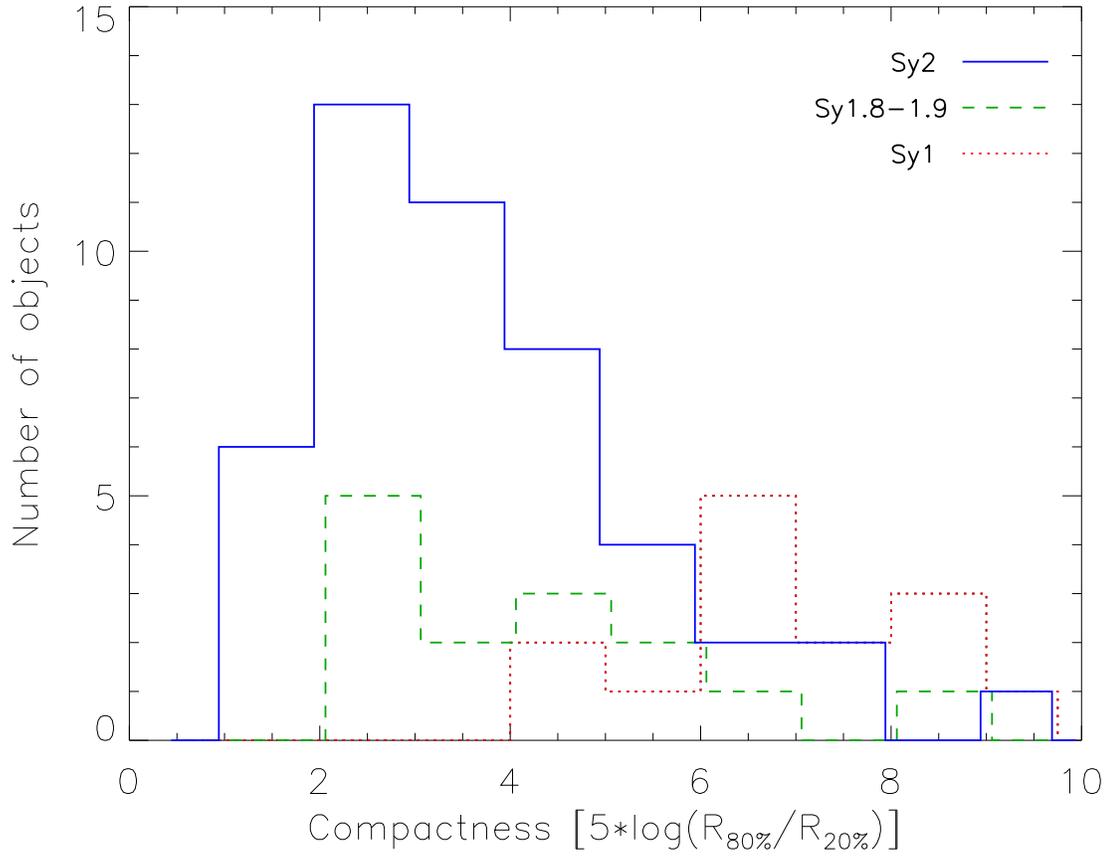}
\caption{Histogram of compactness for Sy subsamples. Type of line and colors codify the Sy type in the same way as in previous figures.}
\label{fig:f14}
\end{figure}

The definition of C is based in the curve of growth and depends on the ratio of two radii enclosing some fraction of the total flux. We have used the formula from Bershady, Jangren \& Conselice (2000),
\begin{displaymath}
C=5\log(r_{80}/r_{20}),
\end{displaymath}
where $r_{80}$ and $r_{20}$ are the radii enclosing $80\%$ and $20\%$ of the total flux within R$_{max}$. Fig.~\ref{fig:f14} shows the distribution of C for the different types of Sy galaxies. The distribution of Sy2 and Sy1 are clearly different, being the Sy1 far more compact. The intermediate types show a behaviour in between the other two subsamples. We find that the values of C for Sy2 are similar to those of local normal galaxies studied in the B band by Bershady et al.~(2000). However, our Sy1 have on average much higher values of C. The occurrence of a compact nucleus in the near-UV is determinant for a high value of C to be measured.

\begin{figure}
\plotone{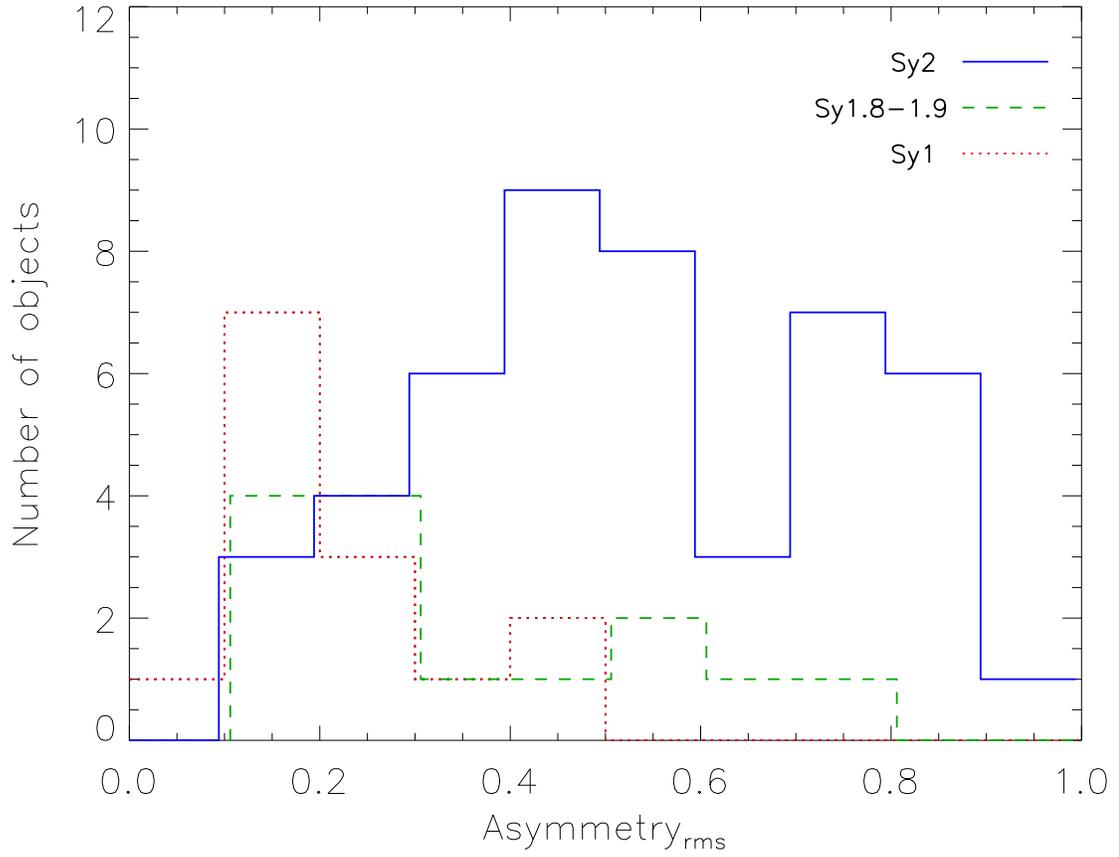}
\caption{Histogram of asymmetry for Sy subsamples. The symbols are the same as in previous figures.}
\label{fig:f15}
\end{figure}

\begin{figure}
\plotone{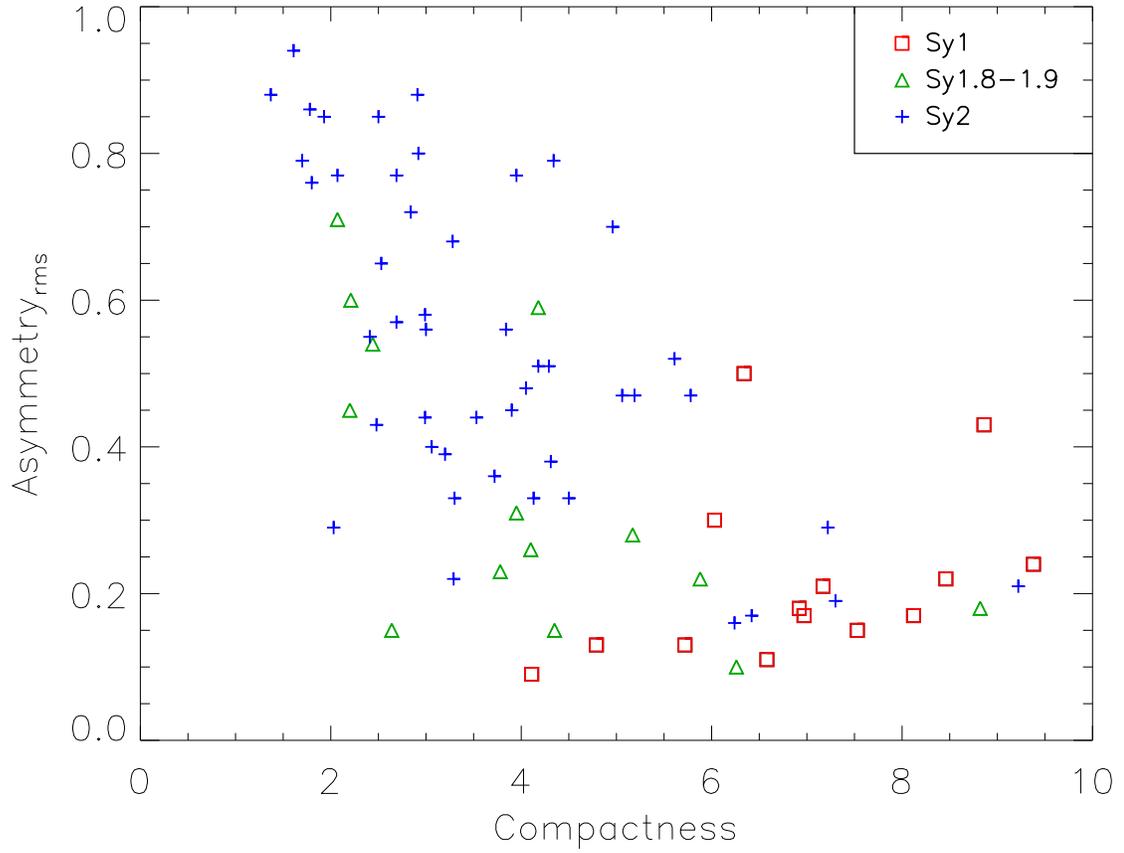}
\caption{Asymmetry vs compactness plot for the galaxies in the sample. Blue crosses stand for Sy2 galaxies, while green triangles represent intermediate types. Sy1 are plotted with red squares.}
\label{fig:f16}
\end{figure}

The asymmetry of a particular galaxy is calculated by subtracting a 180$^\circ$ rotated image from the original one. The residuals in this image are summed up and then normalized dividing by the total flux in the original image. The sum can be quadratic or in absolute value. The rotation center is the nucleus of the galaxy. The formula that summarizes the process is (Conselice 1997)
\begin{displaymath}
A_{rms}^2=\frac{\sum{(F_{ij}-F_{ji})^2}}{2\sum{F_{ij}\,^2}},
\end{displaymath}
where $F_{ji}$ is the rotated original image ($F_{ij}$). A is calculated only taking into account the region inside R$_{max}$, otherwise the contribution of the noise would become important. In the case of M\,81, and M\,51 we have measured only to a radius smaller than R$_{max}$ (5.67$^{\prime\prime}$ for M81 and 6.91$^{\prime\prime}$ for M51), just enough to prevent the occulting 'finger' of the ACS to enter in the region studied, which would introduce a big systematic uncertainty in the asymmetry determination (although is unimportant for the photometry). We decided to use this definition of A$_{rms}$ after trying out as well an absolute value sum (A$_{abs}$). In general, A$_{rms}$ weighs more the bright features, such as star forming regions, and should be less sensitive to the noise. We have checked that the choice of the exact formula does not change the general results, as well as measuring within a half-light radius does not change the general distribution of the points. Figs. \ref{fig:f15} and \ref{fig:f16} show the histogram of A$_{rms}$ values and a A$_{rms}$ vs C plot. The values of A$_{rms}$ for Sy2 are systematically higher, covering a wide range of values, while Sy1 show a very small scatter around A$_{rms}$=0.2. We calculated the asymmetry of some isolated point-like sources leading to a value close to 0.2, so this seems to be a lower limit for the asymmetry calculated by this method. This is a combination of the contribution of the noise and subsampling effects due to the value of the PSF FWHM. Thus, A$_{rms}$ is dominated by the nuclear PSF in Sy1, while extended emission and star-forming regions, together with a smaller nuclear contribution, determines the higher values in Sy2. In Fig.~\ref{fig:f16} it is clearly seen how Sy1.8--1.9 reproduce characteristics of both Sy types 1 and 2. In the plot A$_{rms}$ vs C there is a clear trend that A$_{rms}$ decreases with increasing C, as had been observed before for normal galaxies (see e.g. Bershady et al.~2000). In this plot the correlation saturates when we explore high values of C, due to the limit in A$_{rms}$. Results of the calculation of C and A are summarized in Table~\ref{tab:sample_results}.

\begin{deluxetable}{lcccc}
\tabletypesize{\tiny}
\tablewidth{0pc}
\tablecaption{Results from the shape and stellar clusters analysis.}
\tablehead{
  \colhead{Galaxy}   &
  \colhead{C}   &  
  \colhead{A$_{rms}$}    & 
  \colhead{f$_{clus}$}  &
  \colhead{log(F$_{clus}$)} 
  \\
  \colhead{Name}     &
  \colhead{}   &
  \colhead{}   &
  \colhead{}   &
  \colhead{erg/cm$^2$/s/\AA}  
\\
  \colhead{(1)}     &
  \colhead{(2)}     &
  \colhead{(3)}     &
  \colhead{(4)}     &
  \colhead{(5)} 
  }
\startdata
CGCG\,164-019                   &  6.42    &  0.17  &    0.022  &  -16.69  \\
Circinus                        &   1.7    &  0.79  &    0.016  &  -14.12  \\
ESO\,103-G35                    &  2.41    &  0.55  &        0  &   --     \\
ESO\,137-G34                    &  1.93    &  0.85  &   0.0058  &  -17.12  \\
ESO\,138-G1                     &  3.84    &  0.56  &        0  &   --     \\
ESO\,362-G8                     &  3.53    &  0.44  &        0  &   --     \\
fairall49                       &  2.84    &  0.72  &    0.033  &  -16.31  \\
IC\,2560                        &  5.19    &  0.47  &        0  &   --     \\
IC\,4870\tablenotemark{a}       & $>$9.22  &  0.21  &    0.033  &  -15.71  \\
IC\,5063                        &  2.07    &  0.77  &    0.014  &  -16.62  \\
Mrk\,6\tablenotemark{a}         & $>$5.72  &  0.13  &        0  &   --     \\
Mrk\,40\tablenotemark{a}        & $>$6.97  &  0.17  &        0  &   --     \\
Mrk\,42\tablenotemark{a}        & $>$7.17  &  0.21  &    0.038  &  -16.08  \\
Mrk\,231\tablenotemark{a}       & $>$8.12  &  0.17  &    0.009  &  -16.16  \\
Mrk\,334                        &  5.17    &  0.28  &     0.13  &  -15.45  \\
Mrk\,461                        &   7.3    &  0.19  &   0.0081  &  -16.58  \\
Mrk\,471                        &  2.07    &  0.71  &    0.032  &  -16.45  \\
Mrk\,477                        &  6.24    &  0.16  &        0  &   --     \\
Mrk\,493\tablenotemark{a}       & $>$6.03  &  0.30  &    0.034  &  -15.68  \\
Mrk\,516                        &  4.18    &  0.59  &    0.033  &  -16.78  \\
Mrk\,915\tablenotemark{a}       & $>$6.58  &  0.11  &        0  &   --     \\
Mrk\,1210                       &  7.22    &  0.29  &    0.016  &  -16.40  \\
NGC\,449                        &  5.78    &  0.47  &    0.017  &  -16.81  \\
NGC\,1144                       &  2.91    &  0.88  &    0.032  &  -16.44  \\
NGC\,1320                       &  3.72    &  0.36  &        0  &   --     \\
NGC\,1672                       &  1.61    &  0.94  &     0.13  &  -14.28  \\
NGC\,2639                       &  2.44    &  0.54  &   0.0027  &  -17.02  \\
NGC\,3031\tablenotemark{b,c}    &  2.64    &  0.15  &       0   &   --     \\
NGC\,3081                       &  3.28    &  0.68  &    0.078  &  -15.43  \\
NGC\,3227\tablenotemark{a}      & $>$7.53  &  0.15  &    0.003  &  -16.58  \\
NGC\,3362                       &  2.99    &  0.58  &     0.04  &  -16.45  \\
NGC\,3393                       &  2.48    &  0.43  &        0  &    --    \\
NGC\,3486                       &  3.29    &  0.22  &        0  &    --    \\
NGC\,3516\tablenotemark{a}      & $>$9.38  &  0.24  &        0  &    --    \\
NGC\,3786                       &  5.88    &  0.22  &    0.012  &   -16.74 \\
NGC\,3982                       &  3.06    &  0.40  &    0.031  &   -16.10 \\
NGC\,4253                       &  8.46    &  0.22  &     0.02  &   -16.04 \\
NGC\,4258                       &   2.2    &  0.45  &    0.035  &   -15.18 \\
NGC\,4303                       &  2.03    &  0.29  &     0.13  &   -14.74 \\
NGC\,4395\tablenotemark{a}      & $>$4.35  &  0.15  &   0.0093  &   -17.07 \\
NGC\,4565                       &  2.21    &  0.60  &        0  &    --    \\
NGC\,4593\tablenotemark{a}      & $>$6.34  &  0.50  &   0.0023  &   -16.29 \\
NGC\,4725                       &  2.69    &  0.57  &        0  &    --    \\
NGC\,4939                       &   3      &  0.56  &     0.01  &   -16.95 \\
NGC\,4941                       &  3.2     &  0.39  &   0.0087  &   -16.73 \\
NGC\,5005\tablenotemark{b}      &  2.5     &  0.85  &    0.021  &   -19.05 \\
NGC\,5033                       &  3.78    &  0.23  &    0.007  &   -16.05 \\
NGC\,5135                       &  2.92    &  0.80  &     0.3   &   -14.40 \\
NGC\,5194\tablenotemark{b,c}    &  4.31    &  0.34  &   0.145   &   -14.65 \\
NGC\,5256                       &  5.61    &  0.52  &   0.045   &   -16.23 \\
NGC\,5273\tablenotemark{a}      & $>$8.82  &  0.18  &   0.0056  &  -16.55  \\
NGC\,5283                       &  3.3     &  0.33  &   0.063   &  -16.04  \\
NGC\,5347                       &  4.05    &  0.48  &       0   &   --     \\
NGC\,5548\tablenotemark{a}      & $>$6.92  &  0.18  &   0.0055  &  -16.29  \\
NGC\,5674                       &  3.95    &  0.31  &     0.11  &  -16.04  \\
NGC\,5695                       &  2.99    &  0.44  &   0.039   &  -16.49  \\
NGC\,5728                       &  1.8     &  0.76  &    0.04   &  -15.61  \\
NGC\,5940\tablenotemark{a}      & $>$4.11  &  0.09  &       0   &   --     \\
NGC\,6300                       &  1.78    &  0.86  &      0    &   --     \\
NGC\,6814\tablenotemark{a}      & $>$4.79  &  0.13  &       0   &   --     \\
NGC\,6951                       &  1.37    &  0.88  &    0.13   &  -14.82  \\
NGC\,7130                       &  4.96    &  0.70  &    0.25    &  -14.43 \\
NGC\,7212                       &  4.29    &  0.51  &    0.07    &  -16.17 \\
NGC\,7319                       &  2.53    &  0.65  &  0.023     & -17.64  \\
NGC\,7469\tablenotemark{a}      & $>$8.86  &  0.43  &   0.115    & -14.44  \\
NGC\,7479                       &  2.69    &  0.77  &   0.01     & -17.20  \\
NGC\,7496                       &  3.9     &  0.45  &   0.38     & -14.60  \\
NGC\,7674                       &  5.06    &  0.47  &   0.11     & -15.67  \\
NGC\,7743                       &  4.13    &  0.33  &      0     &  --     \\
UGC\,1214                       &  4.18    &  0.51  &      0     &  --     \\
UGC\,1395                       &  4.1     &  0.26  &       0    &  --     \\
UGC\,2456                       &  4.34    &  0.79  &    0.3     & -15.11  \\
UGC\,6100                       &  3.95    &  0.77  &      0     &  --     \\
UGC\,12138\tablenotemark{a}     & $>$6.26  &  0.10  &       0    &  --     \\
UM\,625                         &  4.5     &  0.33  &    0.1     & -15.90  \\
\enddata
\label{tab:sample_results}
\tablecomments{Col.\ (1): Galaxy name; Col.\ (2): Compactness. Col.\ (3): Asymmetry. Col.\ (4): Fraction of light in stellar clusters. Col.\ (5): Logarithm of the total flux of light in clusters in erg/s/cm$^2$/\AA.}  
\tablenotetext{a}{These objects posses a very bright compact nucleus that affects the determination of R$_{20}$ (see text). In these cases we can only set a lower limit for the compactness parameter.}
\tablenotetext{b}{In these cases R$_{max}$ is limited by the border of the field of view and not by the integration. Compactness and asymmetry are computed based on this smaller R$_{max}$.}
\tablenotetext{c}{Occulting finger of the HRC limits the radius for which the asymmetry parameter (see text) is computed (smaller than R$_{max}$).}
\end{deluxetable}

\clearpage

\subsection{The Fraction of Light in Clusters}

\begin{figure}
\plotone{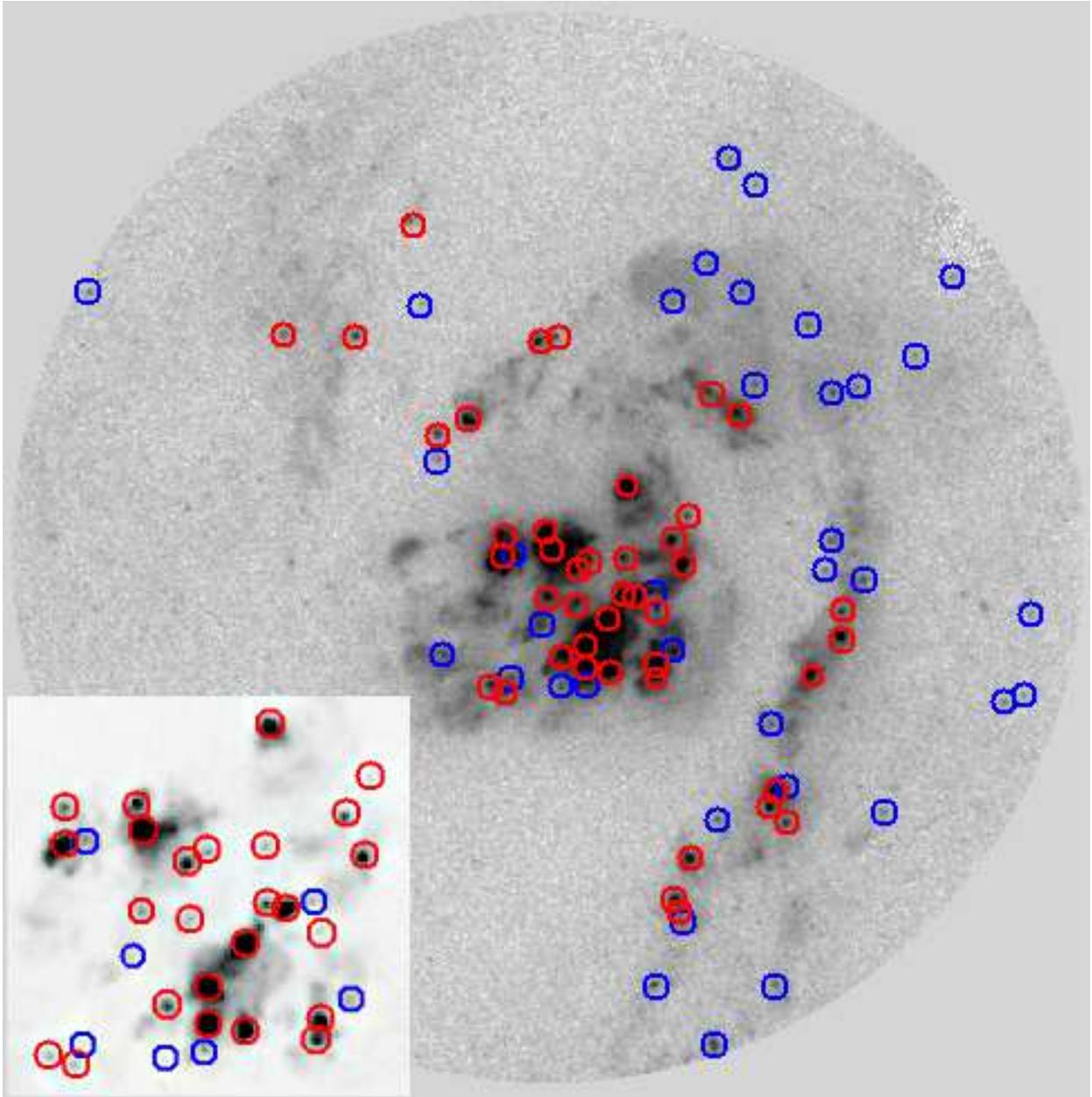}
\caption{An example of the cluster selection for NGC\,5135. Stellar clusters were identified combining a linear display to disentangle the brightest crowded regions (red circles), and then using a logarithmic display, in order to identify the faint population (blue circles). A close-up of the central region is shown in the lower left box.} 
\label{fig:f17}
\end{figure}

In order to determine how important is the contribution of the star formation to the total UV flux we have estimated which fraction of the total flux comes from stellar clusters or very compact light emission (f$_{clus}$), not including the compact Sy1 nuclei. Several software packages were used to detect the clusters. We found that in many cases the highly varying background, and the large dynamical range in some of the images, were a problem for these algorithms to give a satisfying result. Also, the varied morphology became a problem for deciding an homogeneous and unbiased way to perform the analysis automatically. We obtained sometimes good results with IRAF task DAOFIND, but it did not work well in crowded regions or with a highly varying background. Often we had to crop the resulting lists by hand and add some other objects. We thus decided to select the objects by eye inspection. To be sure that we did selections that were complete enough, we checked using linear and logarithmic displays, compared with the optical images in unclear cases, and compared with DAOFIND results. The selections were restricted to the region inside R$_{max}$, in which the total flux was measured, although sometimes there were obvious star clusters outside this region. See Fig.~\ref{fig:f17}, in the electronic edition of the Journal, as an illustrative example. Note that not every clump was added, but only the ones which seemed compact enough to be considered individual clusters or tight aggregations of them. Selecting the clusters by hand proved to be effective, although the limiting magnitude cannot be determined due to the varying background. The completeness of the selection is not critical for this work. Instead, we were interested in checking how robust the estimation of the flux in clusters was with respect to different star cluster selections. Wilson, Harris \& Longden (2005) study the star cluster population in Arp220 with the ACS, finding the same problems. The manual cluster selection also proved to be efficient for them.

The flux determination was done with IRAF task PHOT. We measured the flux within very small apertures, and then we applied aperture corrections from the enclosed energy curves of Sirianni et al.~(1995). Using different apertures led to different results for f$_{clus}$, in part due to the use of a correction for point-like objects, when the star clusters may show a resolved structure, at least for large objects in nearby galaxies. However, in most objects, the variation of f$_{clus}$ when considering different apertures (as 3, 4, 5 or 6 pixels radius) was higher than the variation when using different selection methods. We therefore estimate that the main uncertainty source is the clustering of the objects  and the highly varying background. Finally we decided to use an aperture radius of 4 pixels, as a compromise between the sampling effects of a smaller aperture and the possible aperture overlapping of a larger one, what would also introduce a larger uncertainty in the background subtraction. The background was calculated by measuring in an annulus of 6 pixel of inner radius and 2 pixel width in a median filtered image with a 15$\times$15 pixel box. The fraction f$_{clus}$ was then determined summing up the total flux in all the objects detected inside R$_{max}$, local background subtracted and aperture corrected, and divided between the total flux within R$_{max}$. In Table~\ref{tab:sample_results} we give f$_{clus}$, as well as the logarithm of the total flux in star clusters.

\begin{figure}
\plotone{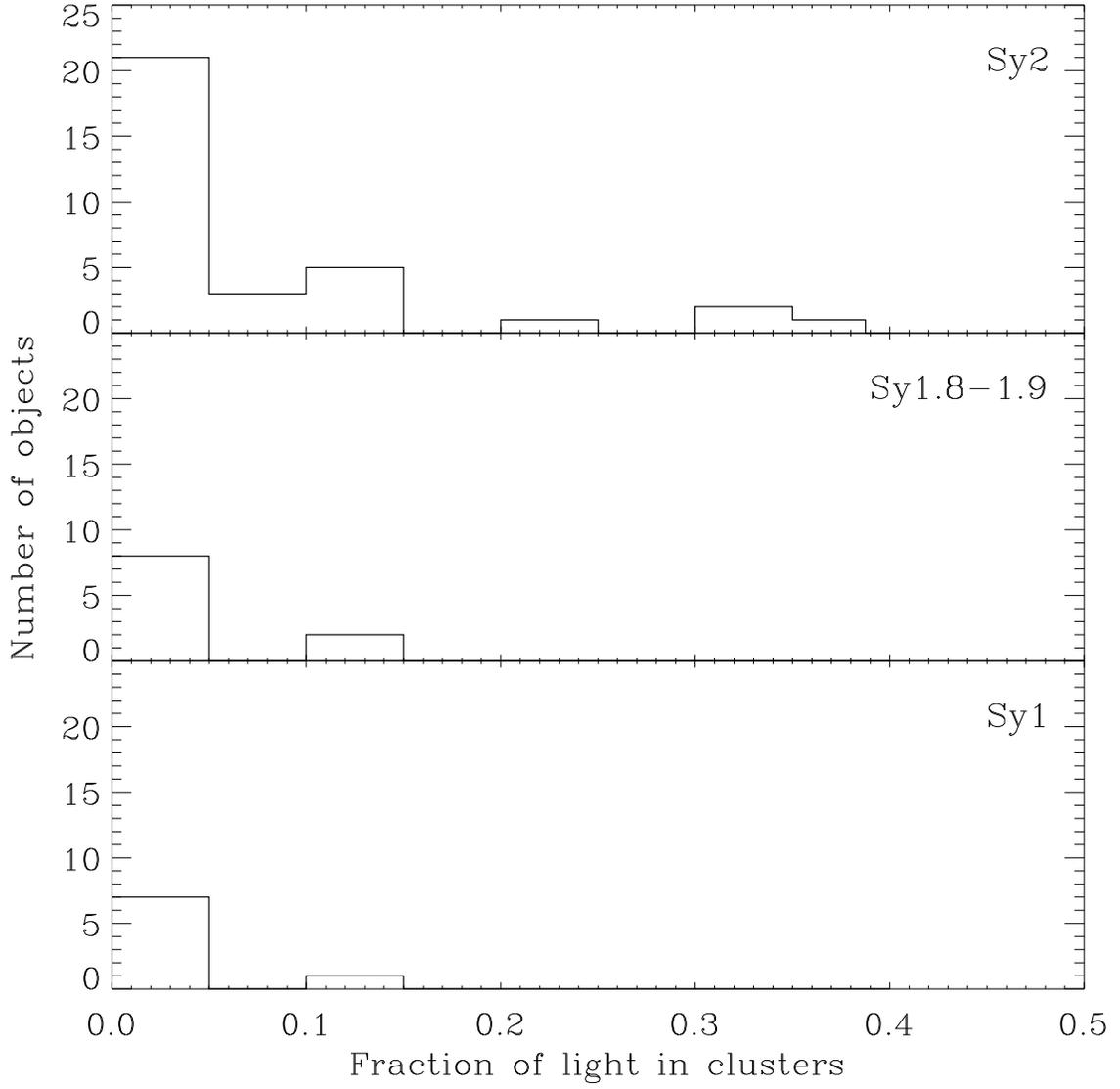}
\caption{Histogram of f$_{clus}$ for different Sy types. Only galaxies with detected star clusters have been included.}
\label{fig:f18}
\end{figure}

In some galaxies we could not detect any star cluster inside the R$_{max}$ aperture. This happens more often for Sy1 (6 out of 14 galaxies; or 43\%) than in Sy2 (14/47; 30\%) or intermediate types (5/14; 36\%). Fig.~\ref{fig:f18} shows a histogram of the distribution of f$_{clus}$ for the galaxies with detected clusters. Except for Mrk\,231, no Sy1 show stellar clusters or star-forming regions contributing more than 5\% to the total flux, while there are 13 ($\sim$ 28\%) of the Sy2 that overcome this value. This confirms that Sy1 galaxies are core dominated objects, while clusters and star-formation account for a significant fraction of the light in Sy2. This fraction is smaller than that calculated for UV-selected starburst galaxies, in which light from clumpy structure is, on average, of the order of 20\% of the total flux (Meurer et al.~1995), although these results are for a different UV filter, at 2200 \AA.

\begin{figure}
\plotone{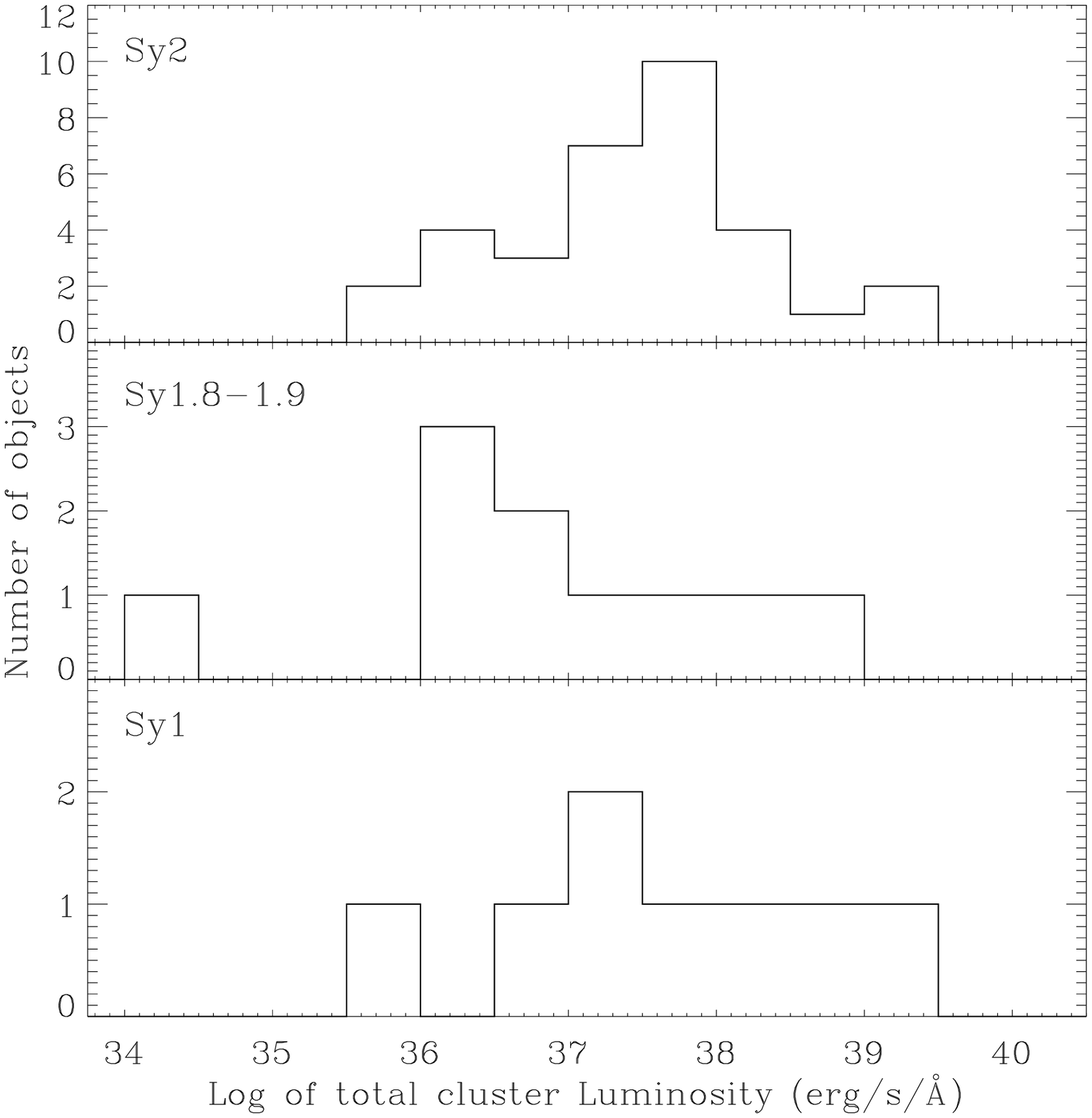}
\caption{Histogram of the total Luminosity coming from star clusters. The three distributions look quite similar.}
\label{fig:f19}
\end{figure}

In order to directly compare the flux coming from clusters among the different subsamples, we plot in Fig.~\ref{fig:f19} a comparative histogram of the total luminosity from stellar clusters. Despite the small number of Sy1 and intermediate type galaxies involved, this Figure shows these sources have a similar distribution of values to that of Sy2. This suggests that these sources have similar amounts of recent star formation, confirming that most of the differences seen in Fig.~\ref{fig:f18} were due to the strong contribution from the nuclear point source in the former.

We have shown that, in spite of the very high resolving power of HST, the bright Sy1 nuclei dominate the emission of the very inner near-UV morphology in this kind of galaxies. In order to unmask possible underlying star-forming regions a very careful PSF-subtracting for Sy1 nuclei is needed. Our team is currently working on this matter, and the results will be presented in a forthcoming paper (Spinelli et al.; in prep.).

\section{CONCLUSIONS}

Using the high resolution of the Advanced Camera for Surveys onboard HST we performed a snapshot survey of a sample of 75 nearby Seyfert galaxies in the near-UV. These observations complete a very useful multi-wavelength database for these AGN, which have also optical and near-IR images available in the HST archive. We have carried out a general analysis of the near-UV images of this sample, consisting in the identification of unresolved compact nuclear sources, extraction of surface brightness profiles, photometry, determination of compactness and asymmetry parameters and identification of the star cluster population. The size of the sample allows us to compare the results of the analysis among different Sy types: Sy1 (including Sy1.2-1.5), Sy2 and intermediate types (Sy1.8-1.9).

The main conclusions from the photometric and morphological study are:

- In general, the morphology in the UV is very irregular, with clumpy and compact structure in most cases.

- Sy1 are completely PSF dominated objects in their inner regions, but Nuker law profiles are detected for some of the galaxies. Inspecting the surface brightness profiles we find 3/14 Sy1 galaxies, and 6/47 Sy2, which posses a star-forming ring. On the contrary, no star-forming rings are found within the intermediate Sy type subsample. Sy2 galaxies present the most varied and irregular profiles. Some profiles follow an exponential, de Vaucouleurs or Nuker law, but most of them cannot be easily classified.
 
- No nucleus is resolved for any of the Sy1 objects, while on the other hand, almost all Sy2 have the nucleus resolved. At least 5/14 Sy1.8-1.9 galaxies show an unresolved compact nucleus.

- In terms of surface brightness at 1$^{\prime\prime}$, and also from the calculation of the integrated magnitude between 0.3$^{\prime\prime}$ and 1$^{\prime\prime}$, we find no significant difference between the host galaxies of Sy1 and Sy2 nuclei. The difference would arise solely due to the presence of the nuclear source in Sy1.

- Sy1 are very compact and have low values of asymmetry, while Sy2 show a very wide range of compactness and asymmetry values.

From the study of the fraction of light in clusters we conclude that:

- Bright star clusters are slightly more often seen in Sy2 ($\sim$70\%) than in Sy1 ($\sim$57\%), or in intermediate Sy types ($\sim$64\%). However, we have shown that this difference may be due to the masking of the inner regions by the contribution of the bright Seyfert 1 nucleus.

- The distribution of the luminosity in star clusters does not change much among different Sy types, when considering only galaxies with detected clusters.

- The contribution of the clusters to the total flux is much more important in Sy2 (where it reaches up to 30\%) than in the other Sy types, but this is at least partially due to the large contribution of the nuclear source in the Sy1 and intermediate Sy types.


\acknowledgements
VMMM research is funded by the Spanish Research Council (CSIC) under the I3P grant program. This work was supported in part by the Spanish Ministerio de Educaci\'on y Ciencia under grant AYA2004-02703. 

\vspace{2cm}
All the figures published only in the electronic edition of the Astronomical Journal will be available to download as EPS files in the URL: \\http://www.iaa.es/$\sim$manuel/publications/paper01.html

\appendix

\section{ATLAS OF THE SAMPLE}
Brief description and notes on individual objects. The descriptions refer to the F330W images unless it is said otherwise.
\subsection{Sy1,1.2 \& 1.5}

\begin{itemize}

\item{\bf Mrk\,6 (Figs.~7.1, 9.1, 20.1):} This object is a Sy1.5 with a very bright saturated core with some extended emission around.

\item{\bf Mrk\,40 (Figs.~7.2, 9.2, 20.2):} This galaxy shows just a plain point-like bright source.

\item{\bf Mrk\,42 (Figs.~7.3, 9.3, 20.3):} A very bright compact nucleus with a very tight wound spiral of star formation of $\sim$300 pc of radius. Many stellar clusters are individually resolved.

\item{\bf Mrk\,231 (Figs.~7.4, 9.4, 20.4):} This is a very powerful galaxy with a bright nucleus and some diffuse circumnuclear emission. There is an arc of star formation about 2 Kpc to the south. This object falls next to the limit of its classification as a Quasar, and is also an IRAS galaxy, very luminous in the FIR (e.g. Soifer, Neugebauer \& Houck, 1987). It shows as well a powerful megamaser emission first detected by Baan (1985).

\item{\bf Mrk\,493 (Figs.~7.5, 9.5, 20.5):} Its morphology resembles that of Mrk\,42, with a very bright nucleus and a tight wound spiral of star formation. However, the distance to this object is 50\% higher than the distance to Mrk\,42, so the individual stellar clusters are poorly resolved.

\item{\bf Mrk\,915 (Figs.~7.6, 9.6, 20.6):} This galaxy is a prototypical Sy1, with a very bright nucleus surrounded by diffuse emission.

\item{\bf NGC\,3227 (Figs.~7.7, 8.1, 9.7, 20.7):} This Sy1.5 shows a bright saturated nucleus and an off-centered bar of star forming regions, which is misaligned with the main galactic bar. This feature is probably caused by a close interaction with the dwarf elliptical NGC\,3226. There is a star forming region 100 pc to the north that is also visible in [OIII] images (Schmitt \& Kinney, 1996). X-ray variability from its nuclear source has also been reported (Gondoin, 2004).

\item{\bf NGC\,3516 (Figs.~7.8, 9.8, 20.8):} There is diffuse light surrounding the bright nucleus of this galaxy up to several hundred pc away. There is also evidence of obscuration by dust to the south and to the north of the nucleus. The northern dusty patches trace a spiral pattern.

\item{\bf NGC\,4253 (Figs.~7.9, 9.9, 20.9):} This is a barred spiral with several bright star-forming knots and star clusters. An important part of the star formation seems to be associated to the  east part of the bar. The bar itself is visible in the UV image.

\item{\bf NGC\,4593 (Figs.~7.10, 9.10, 20.10):} Apart of the bright nucleus, there is a spiral structure of 1 Kpc width with many individually resolved star clusters.

\item{\bf NGC\,5548 (Figs.~7.11, 9.11, 20.11):} This is a face-on spiral with a very bright nucleus. Several hundred parsecs to the north of the nucleus there is an arc of star formation, plus several scattered and relatively isolated stellar clusters still further away.

\item{\bf NGC\,5940 (Figs.~7.12, 9.12, 20.12):} This is a face-on barred spiral with many star clusters and star forming regions tracing the bar and spiral arms. Due to the combination of size and distance, most of this structure is included in the ACS-HRC field of view.

\item{\bf NGC\,6814 (Figs.~7.13, 9.13, 20.13):} In this image there is not much visible apart of the plain PSF of the nucleus and some faint structure of the outer face-on spiral.

\item{\bf NGC\,7469 (Figs.~7.14, 9.14, 20.14):} This is very interesting object, with a faint spiral that becomes a conspicuous ring of star formation in the inner some hundred pc. Many stellar clusters are individually resolved within this region.

\end{itemize}
\subsection{Sy1.8 \& 1.9}

\begin{itemize}

\item{\bf Mrk\,334 (Figs.~7.15, 9.15, 20.15):} This is a peculiar-HII galaxy with irregular nuclear structure, and strong star formation. It is also a strong IR source.

\item{\bf Mrk\,471 (Figs.~7.16, 9.16, 20.16):} It is very obscured by dust. The UV image shows only a point-like nucleus and many scattered star-forming blobs and star clusters. In the optical image the barred spiral structure is better distinguished, with many dust lanes tracing the bar.

\item{\bf Mrk\,516 (Figs.~7.17, 8.2, 9.17, 20.17):} This galaxy is classified as a Sc. Although it looks quite regular in the IR, at the U band is clearly asymmetric, with a star-forming arm to the south that has not counterpart to the north. The nucleus, that looks double in WFPC2 image, is at the limit of the resolution and clearly separated from a bright blob right next to the north ($\sim$100 pc).

\item{\bf NGC\,2639 (Figs.~7.18, 9.18, 20.18):} The nucleus is heavily obscured and no compact source is seen in the images. However, the main spiral structure is visible, with many stellar clusters in the outer region (several Kpc away from the centre).

\item{\bf NGC\,3031 (M81; Figs.~7.19, 9.19, 20.19):} This is the largest galaxy of one of the nearest groups. It is a typical Sa, with a big bulge that fills the whole field of view of the camera. Some dust lanes are seen in the inner region, although no young star clusters are clearly visible in the UV image. Ho, Filippenko \& Sargent (1995) describe it as a LINER.

\item{\bf NGC\,3786 (Figs.~7.20, 9.20, 20.20):} This galaxy shows a nuclear ring of a few hundred pc radius in HST optical images that is incomplete in the UV. An ionization cone coming from the compact nucleus, to the southeast, is clearly detected.

\item{\bf NGC\,4258 (M106; Figs.~7.21, 9.21, 20.21):} The compact nucleus is resolved in our observation. There are many stellar clusters that can be studied individually, and there is a vast amount of absorption by dust in the southwest half of the image. This galaxy hosts a water masing disk that led to the second best determination of the mass of a super-massive black hole (SBH), after the one in the Milky Way (Miyoshi et al.~1995).

\item{\bf NGC\,4395 (Figs.~7.22, 9.22, 20.22):} This is one of those objects for which the nucleus is at the limit of resolution. It is the object of the latest Hubble type in our sample (Sm). With such a small contribution of the bulge, only several scattered stellar clusters are seen apart from the nucleus and a region of diffuse light 10-20 pc to the west of it.

\item{\bf NGC\,4565 (Figs.~7.23, 9.23, 20.23):} This is a nearby edge-on galaxy. The nuclear region is thus very obscured at this wavelength with neat dust absorption. The nucleus appears partially resolved.

\item{\bf NGC\,5033 (Figs.~7.24, 9.24, 20.24):} The galaxy shows an unresolved nucleus together with an ionization cone open to the east (see Mediavilla et al.~2005). Heavy absorption to the west may be responsible for this asymmetry. There is an interesting feature consisting in a bright bar of light coming from the nucleus and extending 2 arcsec to the north. This might be scattered light from the AGN or part of the Extended Narrow Line Region.

\item{\bf NGC\,5273 (Figs.~7.25, 9.25, 20.25):} This is a lenticular galaxy with the typical morphology of an early type galaxy. It shows a point-like nucleus with extended light emission within the central 100 pc. There are some bright areas and dark lanes. The morphology seen with F606W is very similar, suggesting that the dark lanes are caused by thick dust clouds.

\item{\bf NGC\,5674 (Figs.~7.26, 8.3, 9.26, 20.26):} Although classified as barred spiral in the RC3 catalog, this galaxy clearly shows a ring in UV light. The nuclear morphology is very interesting, with several clumps and stellar clusters embedded in a diffuse emission in the central few hundred pc.

\item{\bf UGC\,1395 (Figs.~7.27, 9.27, 20.27):} This object shows a partially resolved nucleus and a circular shell of $\sim$200 pc radius.

\item{\bf UGC\,12138 (Figs.~7.28, 9.28, 20.28):} It shows a bright point-like nucleus and diffuse emission adjacent to the north. To larger scales (several Kpc) it shows a faint filamentary structure.

\end{itemize}
\subsection{Sy2}

\begin{itemize}

\item{\bf CGCG\,164-019 (Figs.~7.29, 9.29, 20.29):} This Sy2 galaxy shows a bright nucleus and a wide open spiral pattern. Some star clusters and knots are visible within the inner Kpc region, as well as diffuse light that might come from an unresolved stellar component.

\item{\bf Circinus (Figs.~7.30, 9.30, 20.30):} This is a nearby spiral with a heavily obscured nucleus. The most prominent feature is a central ring of diffuse light and a star-forming blob 200 pc to the south.
The galactic latitude of this object is very low, so the image may suffer from foreground stars contamination. This galaxy is known to host a nuclear water masing disk in a sub-parsec scale (Greenhill et al.~2003) and to have a kiloparsec scale ionization cone (Marconi et .al. 1994).

\item{\bf ESO\,103-G35 (Figs.~7.31, 9.31, 20.31):} No compact nuclear source is seen in the UV image, but only diffuse light and some blobs. The nuclear region is crossed by dust lanes, what gives it its chaotic structure.

\item{\bf ESO\,137-G34 (Figs.~7.32, 9.32, 20.32):} Neither the UV nor the optical image show an evident nucleus for this object. It has a patchy and chaotic structure with abundant dust lanes and some bright blobs. It is by far the worse contaminated object in the sample by foreground stars, what results evident from the WFPC2 image. With a scale plate of 5 pc\,pixel$^{-1}$ it is difficult to distinguish a star cluster from a foreground star.

\item{\bf ESO\,138-G1 (Figs.~7.33, 9.33, 20.33):} It has a compact nucleus close to our limit of resolution, and a bright asymmetric circumnuclear zone of diffuse light. The east part looks like an ionization cone or scattered light from the AGN.

\item{\bf ESO\,362-G8 (Figs.~7.34, 9.34, 20.34):} This object shows extended light emission around its resolved nucleus, with dusty patches. No stellar clusters or blobs are seen in the circumnuclear region.

\item{\bf Fairall49 (IR1832-594; Figs.~7.35, 9.35, 20.35):} Within the central Kpc, this object shows a wound spiral that ends up in an asymmetric ring of star-forming knots. Several separated star clusters are seen, as well as a bright resolved nucleus. Malkan et al.~(1995) found a non-resolved nuclear source in the IR. We can resolve the nucleus in our UV image. Maiolino \& Rieke (1995), have reclassified it as a Sy1.8, although we have considered here the traditional classification as Sy2.

\item{\bf IC\,2560 (Figs.~7.36, 9.36, 20.36):} A dust spiral is better seen in the optical images. In the UV an irregular extended emission surrounds the resolved nucleus.

\item{\bf IC\,4870 (Figs.~7.37, 9.37, 20.37):} Some extended filaments as well as a lot of faint star clusters are seen in this nucleus. There are also some bright clusters and a very bright point-like source in the center. This object may be in fact an extragalactic HII region with an unusually high ionization lines (Malkan et al.~1998), and the point-like source might actually be a field star.

\item{\bf IC\,5063 (Figs.~7.38, 9.38, 20.38):} It shows very bright compact but resolved blobs within the nuclear region and some bright filaments along the southeast-northwest direction. Those could be scattered light from the AGN.

\item{\bf Mrk\,461 (Figs.~7.39, 9.39, 20.39):} This galaxy shows a resolved nucleus and a faint spiral structure of some Kpc wide.

\item{\bf Mrk\,477 (Figs.~7.40, 9.40, 20.40):} This compact galaxy hosts a very luminous Sy2 nucleus that has been proved by spectropolarimetry to have a hidden Sy1 (Tran, Miller \& Kay, 1992). Heckman et al.~(1997) have shown that it hosts a very compact nuclear starburst. The galaxy is interacting with a companion 50 arcsec to the north. The nucleus is extended and it shows a bright blob close to the northeast. There is as well an arc of star formation further, in the same direction.

\item{\bf Mrk\,1210 (Figs.~7.41, 8.4, 9.41, 20.41):} This is a compact face-on spiral. The tight wound spiral structure is visible in our image more like a ring, as traced by star forming regions. The bright nucleus appears double at close inspection. Tran, Miller \& Kay (1992) showed, by spectropolarimetry, the presence of a hidden BLR.

\item{\bf NGC\,449 (Figs.~7.42, 8.5, 9.42, 20.42):} This object shows a bright resolved nucleus and several stellar cluster and knots. Star forming regions and dust lanes trace a highly inclined spiral.

\item{\bf NGC\,1144 (Figs.~7.43, 9.43, 20.43):} It belongs to an interacting pair of galaxies (NGC\,1143-1144). It shows a very distorted spiral structure with a circumnuclear ring traced out by dust lanes and bright regions. The nucleus is crossed by dark patches of dust.

\item{\bf NGC\,1320 (Figs.~7.44, 9.44, 20.44):} Although resolved, it possesses a bright compact nucleus. Most of the light is confined to a region of less than 100\,pc wide. There is also a remarkable bright and narrow filament extending to the north-west. Dust lanes and extended emission trace a tight spiral patter, although no stellar clusters are clearly detected in our image.

\item{\bf NGC\,1672 (Figs.~7.45, 9.45, 20.45):} This barred spiral harbors a very intense starburst within the inner Kpc. Many bright stellar clusters are individually resolved in the near-UV image. The star formation is mostly arranged in a ring, inside which there is also an extended diffuse emission. The dust distribution seems completely asymmetrical, with heavy absorption to the north-east half of the nuclear region.

\item{\bf NGC\,3081 (Figs.~7.46, 8.5, 9.46, 20.46):} This is a peculiar ringed-galaxy, with two nested rings, the smaller of which is shown in our F330W image. It has a bright resolved compact nucleus, with a bright ionization cone extending to the north. There is as well an important star forming region $\sim$300\,pc to the south-east of the nucleus.

\item{\bf NGC\,3362 (Figs.~7.47, 9.47, 20.47):} Many stellar clusters and star forming regions trace out a wide open spiral pattern. The nucleus is resolved, elongated and have an extension to the west in the form of a bright filament.

\item{\bf NGC\,3393 (Figs.~7.48, 9.48, 20.48):} Kondratko et al.~(2006) have recently detected signatures of a water masing disk in the sub-parsec scale, around the central SMBH. The image shows an s-shaped bright symmetric filament, what seems to be an ionization cone from the central engine.

\item{\bf NGC\,3486 (Figs.~7.49, 9.49, 20.49):} This is a border-line object between Seyfert and LINER, classified as Sy2 by Ho et al.~(1997). It has a bright nucleus and an extended emission with dust patches spiraling inwards.

\item{\bf NGC\,3982 (Figs.~7.50, 9.50, 20.50):} Although this galaxy was classified as ringed in the RC3 catalog, in our Hubble images this feature results clearly identified as a spiral of star-forming regions, star clusters and dust lanes.

\item{\bf NGC\,4303 (M61; Figs.~7.51, 9.51, 20.51):} It has been reclassified as a Low Luminosity AGN, although in the original proposal was included as Sy2. The nucleus is known to host a compact star cluster as the main source of ionizing radiation (Colina et al.~2002). This nucleus is unresolved in our F330W image. It posses a conspicuous star-forming ring at $\sim$250\,pc radius, with many clusters individually resolved.

\item{\bf NGC\,4725 (Figs.~7.52, 9.52, 20.52):} The nuclear morphology of this early type spiral shows not many features apart of the bright resolved nucleus surrounded by an extended emission with a clear exponential profile.

\item{\bf NGC\,4939 (Figs.~7.53, 9.53, 20.53):} The most noticeable feature of this nucleus is a biconical ionization structure coming out from the central source.

\item{\bf NGC\,4941 (Figs.~7.54, 9.54, 20.54):} It shows an extended emission crossed by dark dust lanes. The nucleus has a compact clumpy structure with a bright compact core.

\item{\bf NGC\,5005 (Figs.~7.55, 9.55, 20.55):} The nucleus has both clumpy and diffuse emission, with a very broad dust lane obscuring the north part of the image. An spiral arm is visible to the south with several isolated star clusters and richer star-forming regions.

\item{\bf NGC\,5135 (Figs.~7.56, 8.7, 9.56, 20.56):} This is a nice example of a very strong nuclear starburst, with many bright star clusters individually resolved and two wide open spiral arms, traced by star-forming regions.

\item{\bf NGC\,5194 (M51; Figs.~7.57, 8.8, 9.57, 20.57):} The nucleus is completely obscured and surrounded by a bright extended emission and crossed by dark dust lanes. The inner several hundred parsecs show some few isolated star clusters, while the outer regions are richly crowded with clusters and star-forming regions.

\item{\bf NGC\,5256 (Figs.~7.58, 9.58, 20.58):} This object is in fact a merging system with a double nucleus separated by $\sim$5\,pc. We have studied the northeastern nucleus, that is outstandingly brighter in our near-UV image than its southwestern companion. Actually, this nucleus has been classified as a LINER in the literature (Osterbrock \& Dahari, 1983). The structure of the nucleus is compact, clumpy and irregular.

\item{\bf NGC\,5283 (Figs.~7.59, 9.59, 20.59):} The nucleus is bright and clumpy, with several almost adjacent objects. It shows filamentary structure of gas extending from the nucleus to the northeast.

\item{\bf NGC\,5347 (Figs.~7.60, 9.60, 20.60):} This is a ringed and barred spiral, although in our image only the inner 100\,pc are distinguishable. The nucleus is very bright and conical opening to the northeast. There is also an extended and quite homogeneous emission more obvious to the north side of the nucleus.

\item{\bf NGC\,5695 (Figs.~7.61, 9.61, 20.61):} The nucleus is compact but resolved, with a blob 0.2\,arcsec to the north. There is a faint diffuse emission around it that follows a de Vaucouleurs profile.

\item{\bf NGC\,5728 (Figs.~7.62, 9.62, 20.62):} This is peculiar galaxy with a neatly distorted ring of star-forming regions. The nucleus is completely obscured and it shows an obvious ionization cone opening to the east.

\item{\bf NGC\,6300 (Figs.~7.63, 9.63, 20.63):} This galaxy is heavily obscured, so only a faint diffuse emission can be appreciated in our image.

\item{\bf NGC\,6951 (Figs.~7.64, 9.64, 20.64):} This object shows a very regular ring of star-forming regions and stellar clusters. Inside the ring the surface brightness remains constant. The nucleus is diffuse and extended, with a couple of brighter blobs.

\item{\bf NGC\,7130 (Figs.~7.65, 8.9, 9.65, 20.65):} It shows an interesting morphology, with a ring and an inner bar. It is very rich in star-forming knots and stellar clusters, with a very bright region that is off-center the ring. We have chosen as the galaxy center the centroid of the brightest of these blobs. The morphology of the center in F330W coincides with that of F210M presented in Gonz\'alez Delgado et al.~(1998).

\item{\bf NGC\,7212 (Figs.~7.66, 9.66, 20.66):} This galaxy belongs to a compact group of interacting galaxies. Spectropolarimetric studies have shown the presence of a hidden BLR (Tran, Miller \& Kay, 1992). It shows a clumpy nuclear morphology and irregular diffuse emission.

\item{\bf NGC\,7319 (Figs.~7.67, 9.67, 20.67):} This object has the lowest measured UV flux in the sample. The nucleus is faint and shows an ionization cone opening to the north.

\item{\bf NGC\,7479 (Figs.~7.68, 9.68, 20.68):} The nucleus is small and resolved and there are some scattered star clusters throughout the field of view. It is remarkable a chain of bright clusters 5\,arcsec to the south of the nucleus, that has a north-south alignment.

\item{\bf NGC\,7496 (Figs.~7.69, 9.69, 20.69):} This barred spiral hosts a very powerful starburst in its center. In the image many star clusters can be seen embedded in a diffuse emission. The center has been chosen as the brightest object in the field.

\item{\bf NGC\,7674 (Figs.~7.70, 8.10, 9.70, 20.70):} An obvious spiral structure with many star-forming regions. The nucleus is very bright and embedded in a extended diffuse emission and surrounded by an arc of star formation.

\item{\bf NGC\,7743 (Figs.~7.71, 9.71, 20.71):} This galaxy is classified as an SB, although in our near-UV image no signs of the spiral structure can be detected. The nucleus is bright and it is surrounded by a diffuse emission that appears brighter to the south.

\item{\bf UGC\,1214 (Figs.~7.72, 8.11, 9.72, 20.72):} The nucleus is very bright and is is surrounded by a Kpc-scale symmetrical structure that seems to be an ionization cone.

\item{\bf UGC\,2456 (Figs.~7.73, 8.12, 9.73, 20.73):} It has a bright clumpy nucleus and S-shaped extended emission with three bright stellar clusters in it.

\item{\bf UGC\,6100 (Figs.~7.74, 9.74, 20.74):} Although the SNR in the image is low, a spiral pattern with several star-forming regions is detected. The nucleus is extended and diffuse.

\item{\bf UM\,625 (Figs.~7.75, 9.75, 20.75):} This galaxy has a very bright compact nucleus that is partially resolved in our image. It also possesses a bright star cluster $\sim$150\,pc directly to the west. Apart from this, the emission is diffuse and compact, as most of the light is enclosed within 1\,arcsec from the nucleus.

\end{itemize}

\subsection{Visual Catalogue}
An image-atlas of all the objects is presented. We show the whole field of view of the near-UV images in Figs.~20.1--20.75. In most cases these figures show additional and complementary information to that of Figs.~7 and 8.

\begin{figure}
\figurenum{20}
\includegraphics[angle=270,width=0.9\textwidth]{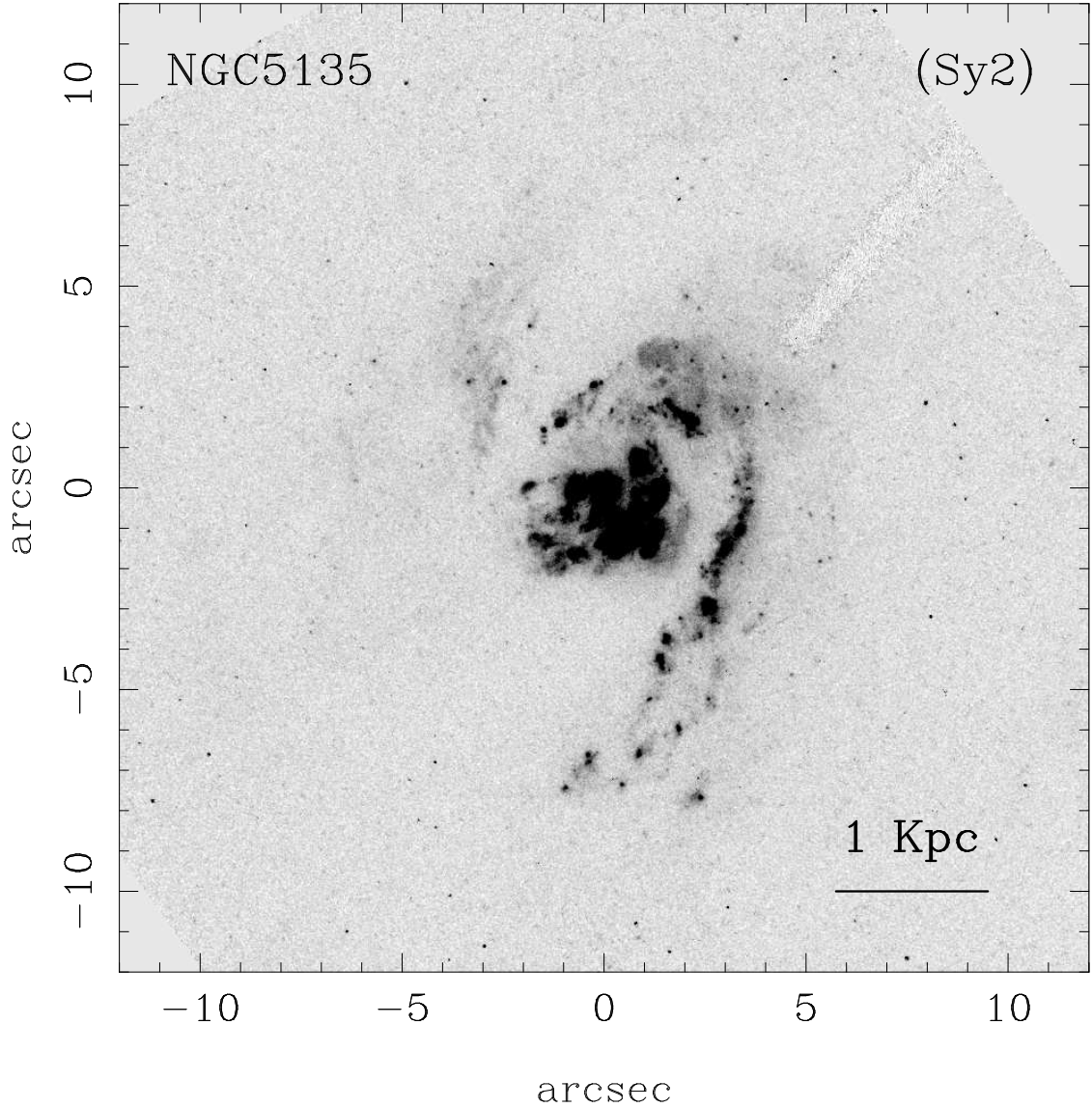}
\caption{ACS example image showing the field of view (FOV) of NGC\,5135. The whole set of FOV images for all the objects in the sample is available in the electronic edition of the Journal. North is up, east to the left.}
\end{figure}





\end{document}